# Modeling the evolution of Gini coefficient for personal incomes in the USA between 1947 and 2005

Ivan O. Kitov

## Introduction

The presence of economic inequality in any modern society is a trivial fact. There are numerous economic theories of income distribution explaining this observation. Neal and Derek (2000) provide a comprehensive overview of the state-of-art in this filed. In spite of the efforts associated with the development of a consistent model of income distribution there are numerous problems yet to resolve. Furthermore, the modern economic theories do not meet some fundamental requirements applied to any scientific theory - a concise description of accurately measured variables and prediction of their evolution beyond the period of currently available measurements.

The most popular aggregate measure of economic inequality is the Gini coefficient. This coefficient is characterized by a number of advantages such as relative simplicity, anonymity, scale independence, and population independence. On the other hand, the Gini coefficient belongs to the group of operational measures: its evolution in time is not theoretically linked to macroeconomic variables and the differences observed between countries are not well explained. These caveats make the Gini coefficient more useful in political and social applications not in economics as a potentially hard science.

As a rule, the Gini coefficient is estimated from household surveys and inequality is reported at family and household level of aggregation. Such an aggregation involves social and demographic processes biasing pure economic mechanisms affecting the inequality. Theoretically, the indivisible level for an inequality study is personal income, which is assumed to be sensitive only to macroeconomic variables. There are just few studies devoted to the Gini coefficient for personal income distribution (PID), however. For example, the US Census Bureau (2006) has been publishing individual Gini coefficients estimated from Current Population Surveys (CPS) since 1994.

Kitov (2005a, 2005b, 2005c, 2006a) has developed a model describing observed personal income distribution in the USA and its evolution through time. This model is based on the prediction of each and every individual income for the population 15 years of age and over. It accurately describes overall PID, the average income dependence on work experience, the evolution of PID in narrow age groups, and the number of people and age dependence in the income zone described by the Pareto distribution. The model also provides predictions for these variables beyond the years where corresponding data are available. Having a complete and precise description of the US PID evolution one can compute relevant Gini coefficients. This makes the Gini coefficient only of secondary importance because its evolution is completely described by the evolution of the PID, which is an exactly modeled function.

The purpose of this study is to accurately estimate the Gini coefficients associated with the personal income distributions provided by the US Census Bureau and to model the evolution of these coefficients between 1947 and 2005, i.e. during the period of continuous PID measurements. For this purpose, an extended analysis of the PIDs has been carried out and the discrepancy between the observed and predicted Gini coefficients is interpreted in the framework of the changing accuracy and methodology, including income definitions, of the CPS during the studied period.

The reminder of the paper is organized as follows. Section 1 introduces the model for the evolution of individual incomes in the USA. Section 2 describes the data on personal income distribution and presents some estimates of Gini coefficients according to various data sets and definitions of income. Section 3 compares the evolution of the observed Gini coefficients with those predicted by the model. Section 4 concludes.

## 1. The model for the evolution of income distribution

The principal assumption of the microeconomic model is that every person above fourteen years of age has a capability to work or earn money using some means, which can be a job, bank interest, stocks, interfamily transfers, etc. An almost complete list of the means is available in the US Census Bureau technical documentation (2002) as the sources of income are included in the survey list. Some principal sources of income are not included, however, what results in the observed discrepancy between aggregate (gross) personal income, GPI, and GDI.

Here we introduce the model described by Kitov (2005a). The rate of income, i.e. the overall income a person earns per unit time, is proportional to her/his capability to earn money, $\sigma$. (An equivalent term for earning money is "work", because work is the only source of any goods and services denominated in monetary units.) The person is not isolated from the surrounding world and the work (money) s/he produces dissipates (conventional economic term for the process would be depreciation, but physical terms are more appropriate in this case) through interaction with the outside world, decreasing the final income rate. The counteraction of external agents, which might be people or any other externalities, determines the price of the goods and services a person creates. The price depends not on some absolute measure of quality of the goods but on the aggregate opinion of the surrounding people on relative merits (expressed in monetary units) of the producers not goods. For example, the magic of famous brands provides a significant increase in incomes for their owners without proportional superiority in quality because people appreciate the creators not goods. As a whole, an equilibrium system of prices arises from the aggregate opinions on relative merits of each and every person not from the physical quantities and qualities of goods and services. The personal incomes are ranked in some fixed hierarchy and, when expressed in monetary units, the hierarchy is transformed in the dynamic system of prices. Since the hierarchy of incomes is fixed, the amounts and qualities of goods can only reorder individuals not change the final aggregate price of everything produced – GDP.

Analogously to many cases observed in natural sciences, the rate of dissipation is proportional to the attained income (per unit time) level and inversely proportional to the size of the means used to earn the money, $\Lambda$. Bulk heating of a body accompanied by cooling through its surface is the case. For a uniform distribution of heating sources, the energy released in the body is proportional to its volume or cube of characteristic linear size and the energy lost through its surface is proportional to the square of the linear size. In relative terms, the energy balance or the ratio of cooling and heating is inversely proportional to the linear size. As a result, a larger body undergoes a faster heating because loses relatively less energy and also reaches a higher equilibrium temperature. Therefore one can write an ordinary differential equation for the changing rate of income earned by a person in the following form:

$$dM(t)/dt= \sigma(t)-\alpha M(t)/\Lambda(t) \qquad (1)$$

where $M(t)$ is the rate of money income denominated in dollars per year [\$/y], $t$ is the work experience expressed in years [y], $\sigma(t)$ is the capability to earn money [\$/y$^2$]; and $\alpha$ is the dissipation coefficient expressed in units [\$/(y$^2$)]. The size of the earning means, $\Lambda$, is also expressed in [\$/y]. The general solution of equation (1), if $\sigma(t)$ and $\Lambda(t)$ are considered to be constant (because these two variables evolve very slowly with time), is as follows:

$$M(t)=(\sigma/\alpha)\Lambda(1-exp(-\alpha t/\Lambda)) \quad (2)$$

In the modeling, we integrate (1) numerically in order to include the effects of the changing $\sigma(t)$ and $\Lambda(t)$. Equations (2) through (4) are derived and discussed in detail below to demonstrate some principal features of the model. These equations represent the solutions of (1) in the case where the observed change in $\sigma(t)$ and $\Lambda(t)$ in all the terms is neglected.

One can introduce the concept of a modified capability to earn money as a dimensionless variable $\Sigma(t)=\sigma(t)/\alpha$. The absolute value of the modified capability, $\Sigma(t)$, and the size of earning means evolves with time as the square root of real GDP per capita:

$$\Sigma(t)= \Sigma(t_0)sqrt(GDP(t)/GDP(t_0))$$

and

$$\Lambda(t)= \Lambda(t_0)sqrt(GDP(t)/GDP(t_0)),$$

where $GDP(t_0)$ and $GDP(t)$ are the per capita values at the start point of the modeling, $t_0$, and at time $t$, respectively. Then the capacity of a "theoretical" person to earn money, defined as $\Sigma(t)\Lambda(t)$, evolves with time as real GDP per capita. Effectively, equation (2) states that the evolution in time of a personal income rate depends only on the personal capability to earn money, the size of the means used to earn money, and the economic growth.

The modified capability to earn money, $\Sigma(t)$, and the size of earning means, $\Lambda(t)$, obviously have positive minimum values among all the persons, $\Sigma_{min}(t)$ and $\Lambda_{min}(t)$, respectively. One can now introduce relative and dimensionless values of the defining variables in the following way: $S(t)=\Sigma(t)/\Sigma_{min}(t)$ and $L(t)=\Lambda(t)/\Lambda_{min}(t)$.

A fundamental assumption is made that the possible relative values of $S(t_0)$ and $L(t_0)$ can be represented as a sequence of integer numbers from 2 to 30, i.e. only 29 different integer values of the relative $S(t_0)$ and $L(t_0)$ are available: $S_1$=2,…, $S_{29}$=30; $L_2$=2,…, $L_{29}$=30. This discrete range results from the calibration process described by Kitov (2005a). The largest possible relative value of $S_{max}=S_{29}$=30=$L_{max}=L_{29}$ is only 15 (=30/2) times larger than the smallest possible $S=S_1$ and $L=L_1$ (in the model, the minimum values $\Lambda_{min}$ and $\Sigma_{min}$ are chosen to be two times smaller than the smallest observed values of $\Lambda_1$ and $\Sigma_1$). Because the absolute values of variables $\Lambda_i$, $\Sigma_i$, $\Lambda_{min}$, and $\Sigma_{min}$ evolve with time according to the same law, the relative and dimensionless variables $L_i(t)$ and $S_i(t)$, $i$=1,…,29, do not change with time retaining the discrete distribution of relative values. This means that the distribution of the relative capability to earn money and the size of earning means is fixed as a whole over calendar years and also over ages. This assumption on the rigid character of the hierarchy of incomes is supported by observations, as presented by Kitov (2005a, 2005b) for the period between 1994 and 2002. This study extends the set of observations to the period between 1947 and 2005.

In equation (2), one can substitute the product of the relative values $S$ and $L$ and the time dependent minimum values $\Lambda_{min}$ and $\Sigma_{min}$ for $\Sigma(t)$ and $\Lambda(t)$. We also normalize the equation to the maximum values $S_{max}$ and $L_{max}$. The normalized equation for the rate of income, $M_{ij}(t)$, for a person with the capability, $S_i$ and the size of earning means, $L_j$ is as follows:

$$M_{ij}(t)/(S_{max}L_{max}) = (\Sigma_{min}\Lambda_{min})(S_i/S_{max})(L_j/L_{max})(1 - exp(-(\alpha/\Lambda_{min}L_{max})t/(L_j/L_{max}))) \quad (3)$$

or

$$M'_{ij}(t) =\Sigma_{min}(t)\Lambda_{min}(t)S'_iL'_j\{1 - exp[-(1/\Lambda_{min})(\alpha' t/L'_j)]\} \quad \textbf{(3')}$$

where $M'_{ij}(t)=M_{ij}(t)/(S_{max}L_{max})$; $S'_i=(S_i/S_{max})$; $L'_j=(L_j/L_{max})$; $\alpha'=\alpha/L_{max}$, $S_{max}$=30, and $L_{max}$=30. Below we omit the prime indices. The term $\Sigma_{min}(t)\Lambda_{min}(t)$ corresponds to the total (cumulative) growth of real GDP per capita from the start point of a personal work experience, $t$ $(t_0=0)$, and is different for different years of birth. This term might be considered as a coefficient defined for every single year of work experience because this is a predefined external variable. Thus, one can always measure the personal income in units $\Sigma_{min}(t_0)\Lambda_{min}(t_0)$**.** Then equation (3') becomes a dimensionless one and the coefficient changes from 1.0 as the real GDP per capita evolves relative to the start year.

Equation (3') represents the rate of income for a person with the defining parameters $S_i$ and $L_j$ at time $t$ relative to the maximum possible personal income rate. The maximum possible income rate is obtained by a person with $S_{29}$=30/30=1 and $L_{29}$=30/30=1 at the same time $t$. The term $1/\Lambda_{min}$ in the exponential term evolves inversely proportional to the square root of real GDP per capita. This is the defining term of the personal income evolution, which accounts for the differences between the start years of work experience. The numerical value of the ratio $\alpha/\Lambda_{min}$ is obtained by calibration for the start year of the modeling. This calibration assumes that $\Lambda_{min}(t_0)=1$ (and $\Sigma_{min}(t_0)=1$ as well) at the start point of the modeling and only the dimensionless factor $\alpha$ has to be empirically determined. In this case, absolute value of $\alpha$ depends on start year.

As numerous observations show, the money earning capacity, $S_iL_j$**,** drops to zero at some critical time, $T_{cr}$, in a personal history (Kitov, 2005a), the solution of (1) is:

$$M_{ij}(t) =$$
$$M_{ij}(T_{cr})exp(-\alpha(t-T_{cr})/\Lambda_{min}L_j) =$$
$$= \{\Sigma_{min}(t)\Lambda_{min}(t)S_iL_j(1-exp(-\alpha T_{cr}/\Lambda_{min}L_j))\}\ exp(-\alpha_1(t-T_{cr})/\Lambda_{min}L_j) \qquad (4)$$

The first term is equal to the level of income rate attained by the person at time $T_{cr}$, and the second term represents an exponential decay of the income rate for work experience above $T_{cr}$. The exponent index $\alpha_1$ is different from $\alpha$ and varies with time. It was found that the exponential decrease of income rate above $T_{cr}$ results in the same relative (as reduced to the maximum income for this calendar year) income rate level at the same age. It means that the index can be obtained according to the following relationship:

$$\alpha_1 = -ln(C)/(A - T_{cr})$$

where $C$ is the constant relative level of income rate at age $A$. Thus, when current age reaches $A$ the maximum possible income rate $M_{ij}$ (for $i$=29 and $j$=29) drops to $C$. Income rates for other values of $i$ and $j$ are defined by (4). For the period between 1994 and 2002, empirical estimates are as follows: $C$=0.72 and $A$=64 years. The observed exponential roll-off for individual and the mean income beyond $T_{cr}$ corresponds to a zero-value work applied to earn money in the model. People do not exercise any effort to produce income starting from some predefined (but growing) point in time, $T_{cr}$, and enjoy exponential decay of their incomes. A physical analog of such decay is cooling of a body, for example, the Earth. When all sources of internal heating (gravitational, rotational, and radioactive decay) disappear, the Earth only will be loosing the internal heat through the surface before reaching an equilibrium temperature with the outer space. This process of cooling is also described by an exponential decay because the heat flux from the Earth is proportional to the difference of the temperatures between the Earth's surface and the outer space.

The probability for a person to get an earning means of relative size $L_j$ is constant over all 29 discrete values of the size. The same is valid for $S_i$, i.e. all people of 15 years of age and above are distributed evenly among the 29 groups for the capability to earn money. Thus, the relative capacity for a person to earn money is distributed over the working age population as the product of independently distributed $S_i$ and $L_j$ - $S_iL_j$ = {2×2/900, 2×3/900, …, 2×30/900, 3×2/900, …, 3×30/900, …, 30×30/900}. There are only 841 (=29x29) values of the normalized capacity available between 4/900 and 900/900. Some of these cases seem to be degenerate (for example, 2x30=3x20=4x15= …= 30x2), but actually all of them define different time histories according to (3'), where $L_j$ is also present in the exponential term. In the model, no individual (in sense of real people) future income trajectory is predefined, but it can only be chosen from the set of the 841 predefined individual futures for each single year of birth.

It is not possible to quantitatively estimate the value of the dissipation factor, $\alpha$, using some independent measurements. Instead, a standard calibration procedure is applied. By definition, the maximum relative value of $L_j$ $(L_{29})$ is equal to 1.0 at the start point of the studied period, $t_0$. The value of $\Lambda_{min}(t_0)$ is also assumed to be 1.0. Thus, one can vary $\alpha$ in order to match predicted and observed PIDs, and the best-fit value of $\alpha$ is used for further predictions. The range of $\alpha/\Lambda_{min}$ from 0.09 to 0.04 approximately corresponds to that obtained in the modeling of the US PIDs during the period between 1960 and 2002 (Kitov, 2005a). Actual initial value of $\alpha$ is found to be 0.086 for $t_0$=1960. The value of $\Lambda_{min}$ changes during this period from 1.0 to 1.49 according to the square root of the real GDP per capita growth. The cumulative growth of the real GDP per capita from 1960 to 2002 is 2.22 times.

Because the exponential term in (2) includes the size of earning means growing as the root square of the real GDP per capita, longer and longer time is necessary for a person with the maximum relative values $S_{29}$ and $L_{29}$ to reach the maximum income rate. There is a critical level of income rate, however, which separates two income zones with different properties. This level is called the Pareto threshold of income. Below this threshold, in sub-critical income zone, PID is accurately predicted by the model for the evolution of individual income. One can crudely approximate the PID by an exponent with a small negative index, as shown later on in the paper. Above the Pareto threshold, in supercritical income zone, PID is governed by a power (equivalent to the Pareto) law. The presence of a high-income zone with the Pareto distribution allows any person reaching the threshold to obtain any income in the distribution, with rapidly decreasing probability, however.

The mechanisms driving the power law distribution and defining the threshold are not well understood not only in economics but in physics as well for similar transitions. The absence of the explicit description of the driving mechanisms does not prohibit using well established empirical properties of the Pareto distribution in the USA – constancy of the exponential index through time and the evolution of the threshold in sync with the cumulative value of the real GDP per capita (Kitov, 2005a, 2005c). Therefore we include the Pareto distribution with empirically determined parameters in our model for the description of the PID above the Pareto threshold. The power law distribution of incomes implies that we do not need to follow each and every individual income as we did in the sub-critical income zone. All we need to know the number of people in the Pareto zone, i.e. the number of people with incomes above the Pareto threshold, as defined by relationships (3) and (4).

The initial dimensionless Pareto threshold is found to be $M_P(t_0)$=0.43 (Kitov, 2005a) and it evolves in time as per capita real GDP:

$$M_P(t)=M_P\,(t_0)(GDP(t)/GDP(t_0)).$$

When a personal income reaches the Pareto threshold, it undergoes a transformation and obtains a new quality to reach any income with a probability described by the power law distribution. This approach is similar to that applied in the modern natural sciences involving self-organized criticality. Due to the exponential (with a small negative index) character of the growth of income rate the number of people with incomes distributed according to the Pareto law is very sensitive to the threshold value, but people with high enough $S_i$ and $L_j$ can eventually reach the threshold and obtain an opportunity to get rich, i.e. to occupy a position at the high-income end of the Pareto distribution.

There is a principal feature of the real PID, which is not described by the model so far, but has an inherent relation to the studied problem. The real income distribution spans the range from \$0 to several hundred million dollars, and the theoretical distribution extends only from \$0 to about \$100,000, i.e. the income interval used in (Kitov, 2005a) to match the observed and predicted distributions. The power law distribution starting from the Pareto threshold income (from \$40,000 to \$60,000 during last fifteen years) describes incomes of about ten per cent of the population. The theoretical threshold of 0.43 was introduced above, partly, in order to match this relative number of people distributed by the Pareto law. The model provides an excellent agreement between the real and theoretical distributions below the Pareto threshold. Above the threshold, the theoretical and real distributions diverge.

Above the Pareto threshold, the model distribution drops with an increasing rate to zero at about \$100,000. This limit corresponds to the absence of the theoretical capacity to earn money, $S_iL_j$, above 1.0. The dimensionless units can be converted into actual 2000 dollars by multiplying factor of \$120,000, i.e. one dimensionless unit costs \$120,000. The observed distribution decays above the Pareto threshold inversely proportional to income in the power of ~3.5. Hence, actual and theoretical absolute income intervals are different above the Pareto threshold and retain the same portion of the total population (~10%). Thus, the total amount of money earned by people in the Pareto distribution income zone, i.e. the sum of all personal incomes, differs in the real and theoretical cases.

Here one can introduce a concept distinguishing below-threshold (subcritical) and above-threshold (supercritical) behaviour of the income earners. Using analogs from statistical physical, Yakovenko (2003) associates the subcritical interval for personal incomes with the Boltzmann-Gibbs law and the extra income in the Pareto zone with the Bose condensate. In the framework of geomechanics as adapted to the modeling of personal income distribution (Kitov, 2005a), one can distinguish between two regimes of tectonic energy release (Rodionov et al., 1982) – slow subcritical dissipation on inhomogenieties of various sizes and fast energy release in earthquakes. The latter process is more efficient in terms of tectonic energy dissipation and the frequency distribution of earthquake sizes also obeys the Pareto power law.

Therefore for personal incomes in the subcritical zone, the income earned by a person is proportional to her/his efforts or capacity $S_iL_j$. In the super-critical zone, a person can earn any amount of money between the Pareto threshold and the highest possible income. A probability to get a given income drops with income according to the Pareto law. The total amount of money earned in the supercritical zone (or extra income) is of 1.33 times larger than the amount that would be earned if incomes were distributed according to the theoretical curve, in which every income is proportional to the capacity. This multiplication factor is very sensitive to the definition of the Pareto threshold. In order to match the theoretical and observed total amount of the money earned in the supercritical zone one has to multiply every theoretical personal income in the zone by a factor of 1.33. This is the last step in equalizing the theoretical and the observed number of people and incomes in both zones: sub- and supercritical. It seems also reasonable to assume that the observed difference in distributions in the zones is reflected by some basic difference in the capability to earn money.

So, the model is finalized. An individual income grows in time according to relationship (3') until some critical age $T_{cr}(t)$. Above $T_{cr}$, an exponential decrease according to (4) is observed. When the income is above the Pareto threshold it gains 33% of its theoretical value (Kitov, 2005b) in order to fit the overall income above the Pareto threshold. Above the Pareto threshold, incomes are distributed according to a power law with an index to be determined empirically. It is obvious that if a personal income has not reached the Pareto threshold before $T_{cr}$, it never reaches the threshold because it starts to exponentially decay. A personal income above the Pareto threshold at critical work experience $T_{cr}$ starts to decrease and can reach the Pareto threshold at some point. Then it loses its extra 33% value.

All people above 14 years of age are divided into 841 groups according to their capacity to earn money. Any new generation has the same distribution of $L_j$ and $S_i$ as the previous one, but different start values of $\Lambda_{min}$ and $\Sigma_{min}$ which evolve with the real GDP per capita. Thus, actual PID depends on the single year of age population distribution. The population age structure is an external parameter evolving according to its own rules. The critical work experience, $T_{cr}(t)$ also grows proportionally to the square root of per capita real GDP. Based on independent measurements of population age distribution and GDP one can model the evolution of the PID below and above the Pareto threshold.

Since the model defines the evolution of all individual incomes in the US economy one can use it for calculation of the Gini coefficient for personal incomes. At the same time, comparison of predicted and measured Gini coefficients obtained for the PIDs is of importance for the model calibration. For example, the Gini coefficient depends on the Pareto law index, $k$, which is also a key parameter of our model.

## 2. The Gini coefficient and the personal income distribution

The Gini coefficient, $G$, is a standard measure of inequality of personal income distribution. By definition, $G$ is the ratio of the area between the Lorenz curve related to a given PID and the uniform (perfect) distribution line, and the area under the uniform distribution line. The Lorenz curve, $Y=F(X)$, is defined as a function of the percentage $Y$ of the total income obtained by the bottom $X$ of people with income. Having measured values of individual incomes for all the population with income and ranking them in increasing order one can precisely calculate corresponding Gini coefficient. It is also possible to include in the consideration those people who do not report nonzero income according to contemporary income definition. In reality, there are some difficulties potentially affecting the accuracy of the PID estimates and the uncertainty of associated Gini coefficients.

The US Census Bureau has been measuring personal income distribution in the USA since 1947 in annual current population surveys. Methodology of the measurements and sample size has been varying with time (US CB, 2002). Therefore, one has to bear in mind potential incompatibility of the CPS results obtained in different years. Changes in income definitions, sample coverage and routine processing influences the estimation of various derivatives of the PIDs, for example, measures of inequality. Moreover, such changes in procedures and definition are likely accompanied by some real changes in true PIDs - the latter changes are hardly distinguished from the former ones. The true PID is the distribution of incomes when all sources of personal income are included.

There are two principal effects of the changing income definitions on the measured PID. First, the number of people with income critically depends on definition of income near zero value. Due to a high concentration of people in the low-income end of the measured PIDs in the USA, the number of people without income is prone to large variations dependent on introduction of new or exclusion of old sources of income in the CPS questionnaires. In addition, it is difficult to give accurate definitions to numerous potential sources of annual incomes near \$1, and even more difficult to distinguish between \$1 and \$2 annual incomes. Due to high uncertainty and low resolution of the current CPS methodology in the low-income end it is practically impossible to measure the true PIDs. Thus, the measured PIDs represent only a varying portion of some true PIDs, the latter being the actual object of our modeling. This situation creates a big challenge for the modeling and interpretation of results.

Figure 1 demonstrates the evolution of a ratio of the number of people with income to the total working age population. There is a significant increase in this ratio: from the lowermost value of 0.64 in 1947 to the highest 0.93 in 1988. The ratio has been slightly decreasing since 1989 - to 0.89 in 2005. Such evolution should be definitely reflected in the estimates of Gini coefficient – people without income introduce a large increase in the coefficient if included. Therefore one has to consider two cases – all population of working age and the portion with income. True PID and Gini coefficient has to be somewhere between these two limit cases. Considering all population, including persons without income according to current definition, one significantly overestimates Gini coefficient, especially in the beginning of the studied period. When only people with income are included, Gini coefficient is underestimated. With time, these two estimates have to converge as the portion of population without income decreases.

Second effect of the changes in income definitions and CPS procedures is related to the change in the portion of aggregate personal income in total GDP. Introduction of new sources of income in the CPS questionnaires should result in an increase in gross personal income, GPI, in addition to the changes in true GPI associated with actual processes. Figure 1 depicts the evolution of the GPI portion in the US GDP: from 0.76 in 1951 to 0.86 in 2001. A severe drop in the portion is observed between 2001 and 2005 – from 0.86 to 0.82. The net change in the GPI portion between 1947 and 2005 is much smaller than the change in the share of the population with income.

A fundamental assumption of the model for the evolution of individual incomes presented in Section 1 is that all people older than 14 years have nonzero annual income and contribute to GPI, which is equivalent to GDI and GDP in our framework. This assumption allows modeling the PID evolution using real GDP per capita, which completely determines time histories of the model defining parameters. The measured PIDs are associated with a changing portion of the GDP.

In addition to the principal difficulties associated with definitions and procedures there are some technical problems for the estimation of Gini coefficient created by the data representation and resolution of the PIDs. The US Census Bureau provides PIDs as the number of people enumerated in income bins of varying width. There were only 14 bins, including the open-end one for very high incomes, in 1947 and 48 bins in 2005. In the absence of information on each and every individual income, Gini coefficient can be calculated by some approximating relation. For example, if $(X_i,Y_i)$ are the values obtained from the CPS, with the $X_i$ indexed in increasing order ($X_{i-1} < X_i$ ), where $X_i$ is the cumulated proportion of the population variable, and $Y_i$ is the cumulated proportion of the income variable, then the Lorenz curve can be approximated on each interval as a straight line between consecutive points and

$$G = 1- \Sigma (X_i-X_{i-1})(Y_{i-1} + Y_i), i=1,\ldots,n \quad (5)$$

is the resulting approximation for $G$. One can also approximate the Lorenz curve using exponential function or power law, where it is appropriate, for interpolation of the underlying PID, as proposed by Dragulesku and Yakovenko (2001).

The choice of the appropriate function for the PID interpolation reveals an important pitfall of the CPS - the usage of the same income bins for representation of counted data during relatively long periods of time. The growth rate of nominal GDP in the USA has been high - more and more people increased their incomes above the upper income limit and found themselves in the group " $MAX and over". So, the coverage of the populations below and above the Pareto threshold, which has been also proportionally growing, differs by several times. This variation in the coverage might potentially result in the increasing or decreasing overall resolution and corresponding bias in the Gini coefficient estimation.

The US Census Bureau (2006) presents several versions of PIDs between 1947 and 2005. In some reports, tables containing PIDs in year specific income bins and counted using current dollars are presented. Some reports give PIDs using CPI-U adjusted (constant) or current dollars but in the same income bins for all the years staring from 1947 to the year of the report issuance. Figure 2 shows some selected original PIDs normalized to the total population (15 years of age and above) for corresponding years and additionally divided by widths of corresponding income bins. Effectively, these curves are probability density functions and show the density of population in $1-wide bin for a given income level. Such a representation allows a direct comparison of the PIDs because they are independent on population size and reduced to the same income bins. As the best approximation, we associate the population density with the mean income in a given bin. These "mean" densities obtained for bins of varying width might be a poor approximation for the densities at the edges of the bins. The wider is the bin the poorer is the approximation. It is worth noting that such a representation excludes the open-end high-income bin because there is no width associated with the bin.

The PIDs between 1947 and 1987 shown in Figure 2a are obtained using same ten income bins as defined by the following boundaries in current dollars: $0, $2000, $4000, $6000, $8000, $10000, $12500, $15000, $20000, $25000, and above $25000. The latter

open-end bin is not shown in the Figure because it does not have finite width for normalization of the PID reading in this bin. Thus only nine bins describe the PIDs between 1947 and 1987.

Figure 2a illustrates the problems with resolution for constant income bins. The PID for 1947 (and also for the years between 1948 and 1950) does not contain any reading for incomes above $9000. This is due to the absence of the persons in corresponding CPS samples reporting such incomes, but not because of the true absence of such people at all. The best resolution (among the PIDs shown in the Figure) at high incomes, i.e. in the Pareto zone, is observed in 1957 – there were seven bins covering the zone. At the same time, there are only two bins covering the low-income zone in 1957. For the PID in 1987, the Pareto threshold is larger than $25000, and the PID contains only one reading in the Pareto zone corresponding to the open-end bin, i.e. the reading not shown in the Figure. As expected, this PID provides the best resolution in the low-income portion of the distribution – nine bins. Therefore the constant bins fail to provide a uniform description of the PIDs between 1947 and 1987. As a result, the estimation of Gini coefficient can be severely biased.

The PIDs between 1947 and 2005 presented in Figure 2b are characterized by income bins which are better adjusted to the observed PIDs (US CB, 2007). These bins cover better than in Figure 2a both low and high incomes, also with varying resolution, however. The years after 1994 are characterized by the highest resolution, i.e. the narrowest income bins of $2500 between $0 and $10000. Because of the increasing number of people with incomes over $100,000, three $50000-wide bins were introduced in 2000, covering incomes up to $250,000, extra to those provided by standard CPS reports. These wide bins provide a more accurate representation of the Pareto distribution and corresponding Gini coefficient.

Kitov (2005b) found that the PIDs between 1994 and 2002 practically collapse to one curve, when normalized to the total working age population and nominal GDP per capita – in other words to nominal GDP. (The normalized PIDs effectively represent portion of population as a function of portion of total income.) This observation demonstrates a fundamental property of the personal income distribution in the USA – it is characterized by a fixed hierarchy of incomes, which changes slowly in time according to the evolution of age structure.

The PIDs measured for the years before 1994 allow to validate this property and to extend the presence of such a fixed hierarchy in the PIDs by 47 years back in the past and 3 years ahead. There is a problem related to the normalization factor, however. The years between 1994 and 2002 are characterized by constant values of the portions of the GPI in the GDP and the population with income in the total working age population, as Figure 1 demonstrates. This means that the nominal GDP grows in sync with the nominal GPI. This is not the case for the years before 1980, however. Therefore one has to replace the nominal GDP with nominal GPI in order to accurately represent the evolution of the PIDs after 1947. Such a procedure has to compensate the difference in the evolution of the GDP and GPI because less sources of personal income were considered in earlier years and income scale was effectively biased down. Figure 3 displays the cumulative growth of the nominal GDP and nominal GPI between 1960 and 2005 reduced to total population and population with income. The curves diverge with time what allows a more robust choice of an appropriate variable for a normalization converting all the observed PIDs into one curve. Figure 4 depicts the PID for 2005 normalized to the four variables in Figure 3. One can clearly distinguish between the resulting normalized PIDs in the low- and high-income zones.

Figures 5 and 6 present results of the normalization of the measured PIDs to the nominal GPI, as reduced to the people with income. For the period between 1947 and 1987 (Figure 5), where PID is measured in the same bin set, the normalized PIDs practically collapse to one curve with only minor deviations probably associated with measurement errors. For the period between 1947 and 2005, where a higher resolution with varying width of income bin is available, the normalized PIDs for population with income (Figure 6a) are also very close. The narrower bins result in higher fluctuations due to measurement errors, however. At the same time, the normalized PIDs for all population of 15 years of age and over demonstrate a larger divergence with time because the normalization is associated with the nominal GPI reduced to the population with income.

The normalized PIDs in Figures 5 and 6a are very close. This observation extends the presence of the fixed hierarchy of incomes, as expressed by the portion of population having given portion of total income, to the years between 1947 and 1993, and beyond 2002. Therefore, one can expect only a slight variation in Gini coefficient related to the PIDs. The presence of the hierarchy also represents a strong argument in favor of our model for the evolution of individual and aggregate income.

Having studied some principal properties of the PIDs for the years between 1947 and 2005, one can start a direct estimation of Gini coefficient using the approximation presented in (5). There are several important problems related to the discrete representation to be resolved, however. The PIDs provide only estimates of aggregate population but not total income in the bins. For the years after 2000, mean income is given for each bin allowing for an accurate estimate of cumulative income. Mean incomes are not reported for the previous years, however.

If to replace mean incomes with central points of corresponding bins, one obtains a slightly biased value of Gini coefficient, as Figure 7 shows. Thus we need a more reliable estimate of the mean income than provided by the central points. The best choice would be to approximate the observed PIDs in the low-income zone, i.e. below the Pareto threshold, by an exponential function, to determine corresponding index for each year, and to precisely calculate Gini coefficient for the given approximation. This approach might potentially provide the most accurate estimate of the PIDs and Gini coefficient if corresponding population estimates in the bins are accurate. Unfortunately, the accuracy is inhomogeneous over the bins of varying width and the advantages of the exponential approximation may disappear, as Figure 8 demonstrates. Therefore, we use a different approach in the low-income zone.

Mean income estimates are available between 2000 and 2005 and it is easy find their average distance from the central points of relevant bins. Figure 9 presents such deviations and corresponding regression line (mean deviation) for 2001 and 2005. The average dimensionless distance, i.e. the difference in \$ divided by the bin width in \$ (\$2500 for the years between 2000 and 2005), is -0.12. Thus, in the following estimation of Gini coefficient we use the mean income values corrected for this deviation in the low-income zone.

In the high-income zone, a power low approximation is a natural choice for the PIDs, as demonstrated in Figures 5 and 6. Theoretically, the cumulative distribution function, *CDF*, of a Pareto distribution is defined by the following relationship:

$$CDF(x) = 1 - (x_m /x)^k$$

for all $x>x_m$, where $k$ is the Pareto index. Then, probability density function, *pdf*, is defined as

$$pdf(x) = kx_m^k/x^{k+1} \qquad (6)$$

The functional dependence of the probability density function on income allows an exact calculation of total population in any income bin, total and average income in this bin, and the input of the bin to corresponding Gini coefficient because the pdf exactly defines the Lorenz curve. Thus, if populations are counted in some predefined income bin set then relevant Lorenz curve can be retrieved using a known value of the Pareto index $k$. We use (6) in the following calculations of empirical Gini coefficients in the Pareto zone. By definition, the Pareto threshold evolves proportionally to the nominal GPI per capita, as described above. Such evolution provides the unchanged shape of the normalized PIDs because it retains unchanged the income value where the transition from the low- to high-income zone occurs.

Now we are ready to estimate Gini coefficients from the measured PIDs using corrected average incomes in the low-income zone and power law approximation in the high-income zone, the transition point evolving proportionally to the nominal GPI per capita. To begin with, we compare our estimates of $G$ with those reported by the US Census Bureau, as shown in Figure 10. For the years between 1994 and 1997, the curves are very close. In 1998, a sudden drop by ~0.01 in the CB curve is not reproduced by the estimated one. There is no apparent reason for the drop – macroeconomic or related to the CPS procedures. It is likely that there was some change in the US Census Bureau approach to Gini coefficient calculation in 1998. After 1998, the curves continue to slightly diverge but move in sync otherwise. The difference between the curves reaches 0.01 in 2005. Our estimates of $G$ seem more consistent and used in further comparisons.

Figure 11 presents estimates of $G$ for the PIDs in current dollars which do not include people without income. There are "crude" estimates of $G$ obtained for the populations counted in the same income bins between 1947 and 1987. A "fine" PID is available from the year specific bins for the period between 1947 and 2005. In spite corresponding income bins were used several years, the overall resolution of the PIDs is higher and $G$ estimates are potentially of lower uncertainty. Two curves in the lower panel of Figure 11 present the evolution of $G$ for the two sets of PIDs – crude and fine ones. In 1947, the curves are spaced by 0.1. When approaching 1970, they slowly converge. Between 1974 and 1984, the curves are hardly distinguished. In 1985, a new period of divergence starts. The observed discrepancy between the curves is related to the coverage of the PIDs by corresponding sets of income bins.

The upper panel of Figure 11 illustrates the difference showing two Lorenz curves for 1947. The crude set of bins does not resolve Lorenz curve well and relevant Gini coefficient is highly underestimated as compared to the fine set. For the years between 1974 and 1984, both sets provide a compatible resolution (the number of bins is 10 and 18, respectively) and the estimates of $G$ converge.

In fact, the Gini curve associated with the fine PIDs is a constant near 0.51 between 1960 and 2005 despite a significant increase in the GPI/GDP ratio and the portion of people with income during this period (see Figure 1). This is a crucial observation because of the famous discussion on the increasing inequality in the USA as presented by the Gini coefficient for households (US CB, 2000). Obviously, the increasing $G$ for households reflects some changes in their composition, i.e. social processes, but not economic processes as defined by distribution of personal incomes.

Between 1947 and 1960, the fine $G$ curve monotonically grows from 0.45 to 0.50. This growth may be associated with the increasing resolution in corresponding PIDs. One can expect a further increase in the estimates of Gini coefficient when a finer grid is used. Possibility of a slight increase in the $G$ estimates associated with the inclusion of new (and true) income sources is not excluded. All in all, Gini coefficient for true PIDs is likely to be higher than that predicted using fine PIDs for the population with income.

In the absence of the true PIDs one can carry out an estimate for a limit case – to include people without income in the first bin of zero width. It is hard to believe that such people might potentially survive, but definitions of the US CB do not allow including income sources of the types corresponding to the incomes of people without income. In any case, the inclusion of these people in the PIDs creates a problem for the estimation of Gini coefficient. Figure 12 presents two time series of $G$ estimates for the crude and fine bin sets. In 1947, the difference is 0.04 what can be explained by the properties of corresponding Lorenz curves, as Figure 12a depicts. Then the curves converge and intercept in 1971. Between 1971 and 1984, the curves are very close and diverge again since 1985. These observations are similar to those associated with the PIDs for the population with income. The only difference is that the curves for the PIDs with total population undergo an expected decrease with time according to the decreasing portion of the population without income. Therefore, the Gini coefficient curves associated with the total working age population and with its portion having nonzero income converge. When the portion with income will reach 1.0, i.e. everybody will have has a nonzero income, the curves become identical. So, where is our prediction relative to the empirical curves?

## 3. Comparison of observed and predicted Gini indices

In the model, the evolution of personal incomes is defined by a number of parameters which have to be determined empirically. For the estimation of Gini coefficient a critical value is the Pareto law index, $k$, which defines how "thick" is the PID tail in the high-income zone. There are two independent techniques for the estimation of $k$.

First, for a Pareto distribution with index $k$ and the minimum value $x_m$, the mean value is

$$x_{av}= (k+1)x_m/k.$$

Therefore, the measured average incomes for the open-end income bins provide valuable information on corresponding Pareto indices: $k = x_m/(x_{av} - x_m)$. Figure 13 presents the estimates of index $k$ for the years between 2000 and 2005, where the average incomes are available. This allows multiple estimates using the increasing number of people with incomes above \$250,000, \$200,000, \$150,000, and \$100,000 – all in the Pareto zone. For example, the average income for people with incomes above \$100,000 in 2005 is \$176,068 and $x_{av}$ for people with incomes over \$250,000 is \$470,616. Corresponding Pareto index values are 1.31 and 1.13. The latter estimate is obtained using the average incomes for male and female separately, as presented by the Census Bureau. There were 10,896,000 people with income above \$100,000 and only 1,334,000 above \$250,000. Bearing in mind that the population estimates are obtained using only a relatively small sample and population controls, one can consider the Pareto index for the population with incomes over \$250,000 as less reliable than the former value. Also, the average income in the open-end bin may be slightly shifted up due to the effect of super-rich people, who do not obey the Pareto distribution. Such a deviation from a power law distribution is also often in hard sciences and usually considered as statistical fluctuation related to the under-representation in the overall population. For example, catastrophic earthquakes do not obey Guttenberg-Richter frequency-magnitude relation. According to Figure 13, $k$=1.3 is our best choice.

Second method is a direct estimation of $k$ using regression in the Pareto income zone. For such a regression we represent the PIDs in log-log scale as shown in Figure 14. The power law index obtained by a linear regression is -3.35. The Pareto index is 1.35 because we use population density income distributions, i.e. the original PIDs reduced to width of corresponding income bins. This normalization effectively increases the power law index in the PIDs by one unit. Therefore, the Pareto index is $k$ = 3.35-2 = 1.35 and consistent with the results obtained by the first method.

Figure 15 demonstrates the effect of $k$ on the predicted Gini coefficient. Obviously, lower $k$ values create "thicker" tails in corresponding PIDs and larger $G$ values. The effect of $k$ on $G$ is nonlinear and the difference of 0.3 units in the index value results in Gini coefficient difference of 0.01 to 0.015. One should not neglect such a difference when comparing predicted and measured Gini indices.

Another parameter of the model which critically depends on the Pareto index is the effective increase in income production in the model relative to that in the sub-Pareto income zone (see Section 1 for details). Figure 16 depicts the dependence of the corresponding ratio on *k*. Empirical value obtained in previous studies (Kitov, 2005a, 2005c) is 1.33. This value corresponds to *k*=1.35. This ratio is very sensitive to *k* and the effect is also slightly nonlinear.

Having estimated the empirical parameters defining the model and the age structure of the US population between 1947 and 2005 (US CB, 2007) one can predict Gini coefficient (for personal incomes) during the studied period. Figure 17 compares the measured and predicted Gini coefficients. The predicted curve is in a good agreement with that obtained using the PIDs for the persons with income. The latter curve lies below the former one, however, during the entire period. The empirical Gini coefficient for the PIDs including all working age population is always above the predicted curve. Therefore the predicted curve takes the place just between the empirical ones and the latter two likely to converge to the predicted one in future when adequate definitions of income will be introduced.

This is an expected result of the modeling – neither of the US Census Bureau income definitions provides an adequate description of the personal income distribution and thus fails to predict Gini coefficient. Usage of such biased *G* values may lead to economic misinterpretation and social confusion. The Gini coefficient for personal incomes in the USA underwent a slight increase between 1947 (0.5346) and 1962 (0.5378), and then has been monotonically decreasing to the current value of 0.524. There was no significant increase in the economic inequality in the USA during the last 60 years as expressed by the Gini coefficient for personal incomes.

## 4. Conclusion

There are several simple but meaningful findings related to the estimation of the empirical Gini coefficients. First and most important consists in the fact that the estimates of Gini coefficient critically depend on definition of income. The inclusion of new income sources in the CPS questionnaires has resulted in a large change in the number of people with income and also in the ratio of GPI and GDP. The current set of definitions is far from that necessary for measuring true PIDs, however.

Second, the Gini coefficient associated with whole population 15 years of age and over and that associated with people with nonzero income converge with time as the portion of people without income decreases. The true Gini coefficient had to be somewhere between these two estimates. Thus the empirical estimates can not be considered as reliable for the purposes of economics as a theory.

Third, the resolution of the empirical PIDs, i.e. the coverage of population with income bins in some proportional ways, influences the estimates of Gini coefficient. A higher resolution guarantees a smaller variation in Gini coefficient over time. Poor resolution leads to a negative bias in the $G$ estimates.

Fourth, the empirical PIDs collapse to practically one curve when normalized to the cumulative growth in nominal GPI for the studied period between 1947 and 2005. The remaining differences in the PIDs are well reflected in the changes of the Gini coefficient obtained using the population with income. The empirical PIDs support the assumptions of the model for the evolution of individual income.

The model predicts practically unchanged (normalized) PIDs and Gini coefficient between 1947 and 2005. Slight changes in the PIDs and $G$ are related to economic growth and changes in the age structure of American population. The decreasing portion of young and thus relatively low-paid people in the working age population effectively leads to a decrease in Gini coefficient. The increasing portion of the population older than the critical age $T_{cr}$ (55 years in 2005) results in an increase in the portion of relatively poor people because of the exponential decrease of income (including average one) with age, as described in Section 1. As a net result of the two effects, the empirical Gini coefficient has a minimum of 0.5238 in 1990 and then starts to grow again, reaching 0.5266 in 2005.

Such defining model parameters as the Pareto law index (1.35) and the ratio of the efficiency of money earning in the Pareto zone relative to the predicted by the model (1.33) are well calibrated by the empirical PIDs and Gini coefficient. The model for the individual income evolution is very sensitive to these parameters.

The empirical Gini curves converge to the predicted one when the number of people without income according to currently adopted definitions decreases. Asymptotically, the empirical curves should collapse to the theoretical one when all the working age population will obtained an appropriate definition of their incomes. Such convergence should be seen clearly in age dependent PIDs, where the portion of the population without income decreases with age. For example, in the age group between 45 and 54 years this portion increased from 0.78 in 1960 to 0.94 in 2005. Hence, the portion was consistently larger and changed less than that for the total working age population (see Figure 1). One can expect a lower difference between the two empirical estimates and a better prediction. This is a subject of a paper in preparation.

Gini coefficient is a crude and secondary measure of inequality for economics as a science. It could be useful for social and political discussions as a relative and operational measure without any specific meaning of its absolute value. What is important and has a primary significance for scientific models are the PIDs, which demonstrate a fixed hierarchy during a very long period between 1947 and 2005. (It is very unlikely that this hierarchy will be destroyed in near future.) The shape and the evolution of the measured PIDs is well predicted for the whole period between 1947 and 2005. This allows exact prediction of the Gini coefficient and other measures of inequality, which are defined using personal income distribution.

The Census Bureau focuses its attention on Gini coefficients related to the measurements of income inequality at family and household level. Corresponding coefficients change with time and are presented as evidence in favor of an increasing economic inequality in the USA (US CB, 2000). Our estimates of the Gini coefficient for the PIDs, both empirical and theoretical, demonstrate that the inequality is not changing so dramatically. Therefore the Gini coefficient associated with households should be affected primarily by some changes in their structure but not by the changes in the personal incomes of the people constituting the household.

## Figures

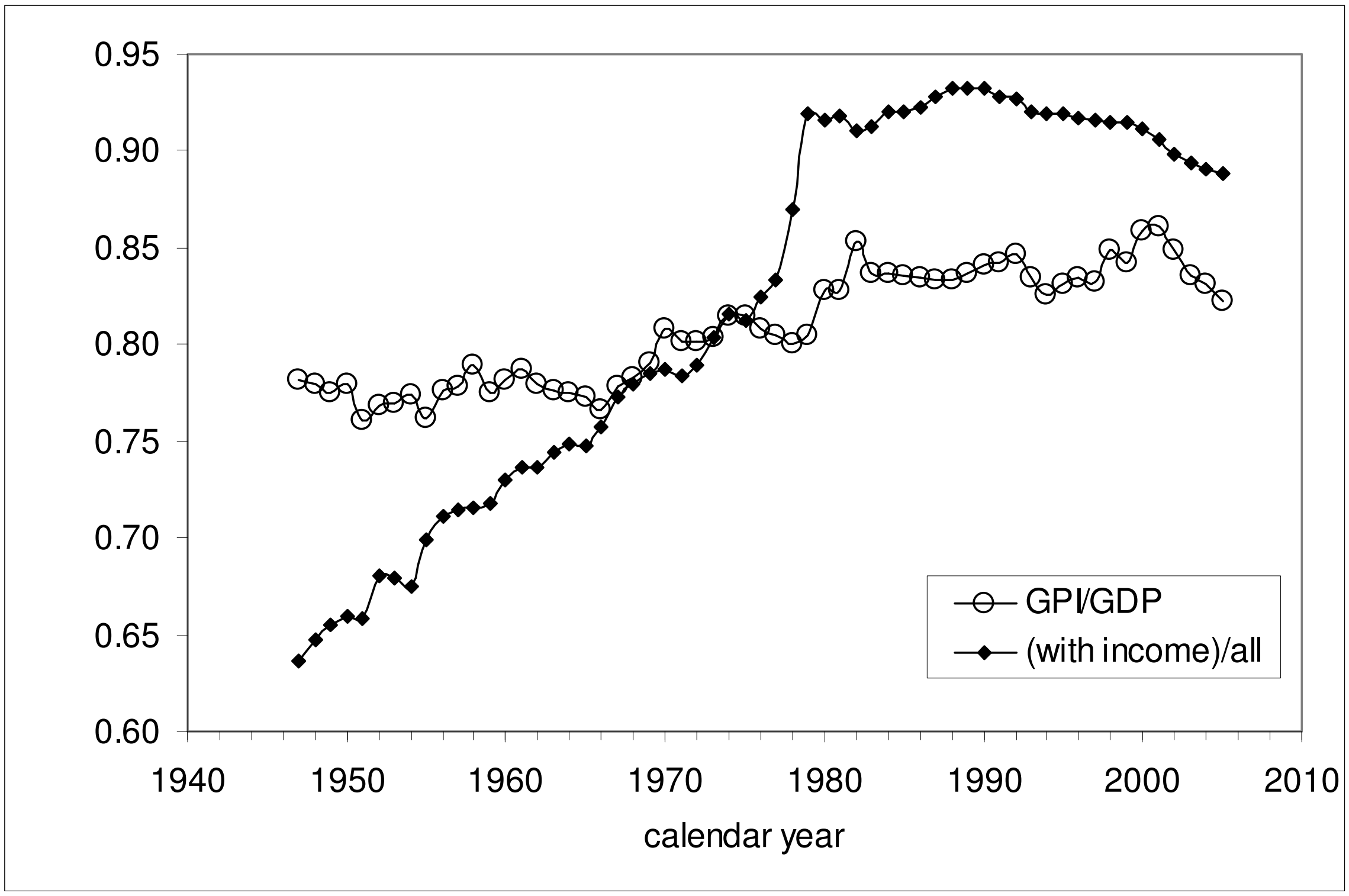

Figure 1. Ratio of the nominal gross personal income, GPI, and nominal GDP for the same year, and ratio of the US population with income, as reported by the US Census Bureau, to the total population of 15 years of age and over.

a)

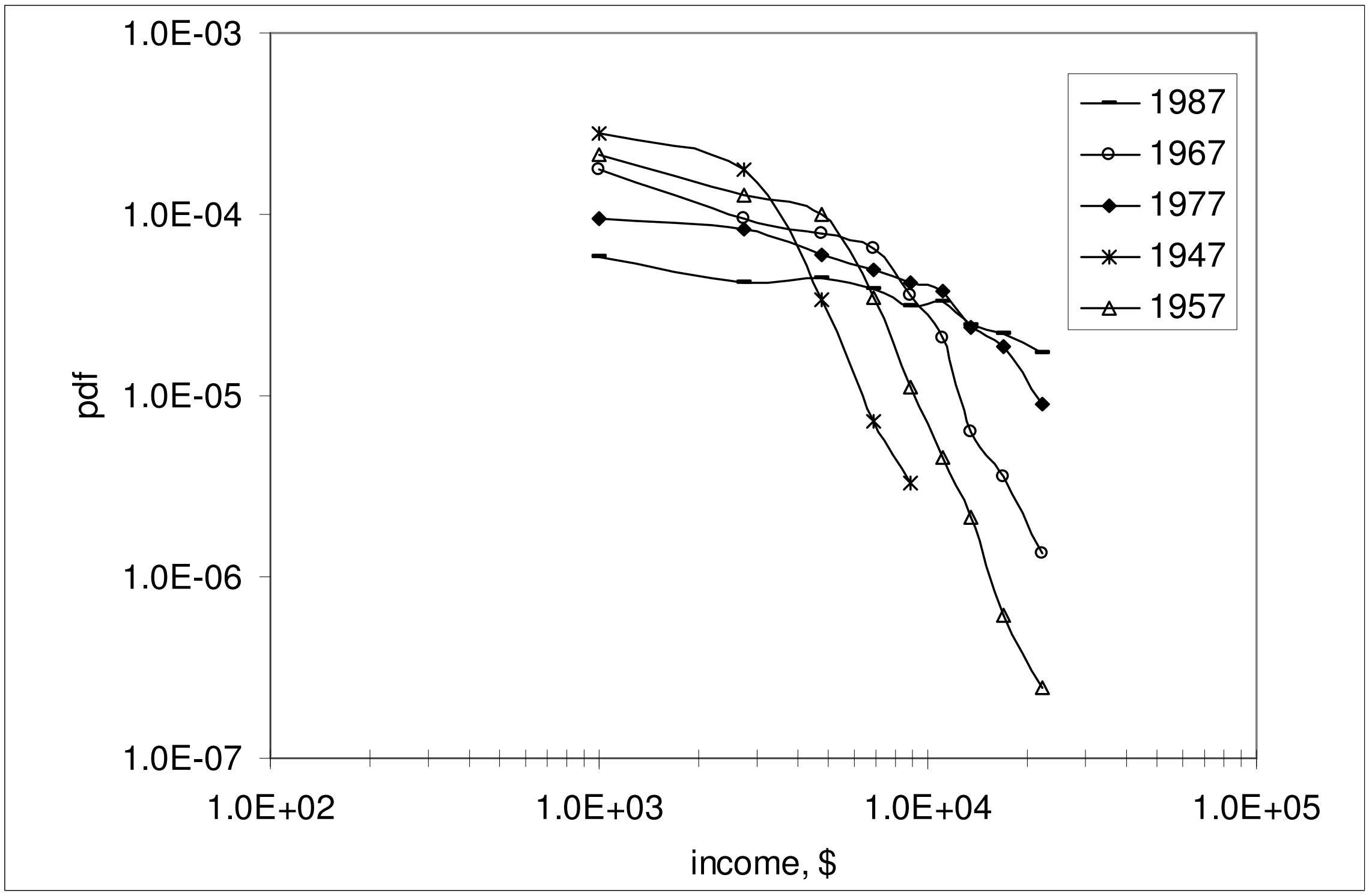

b)

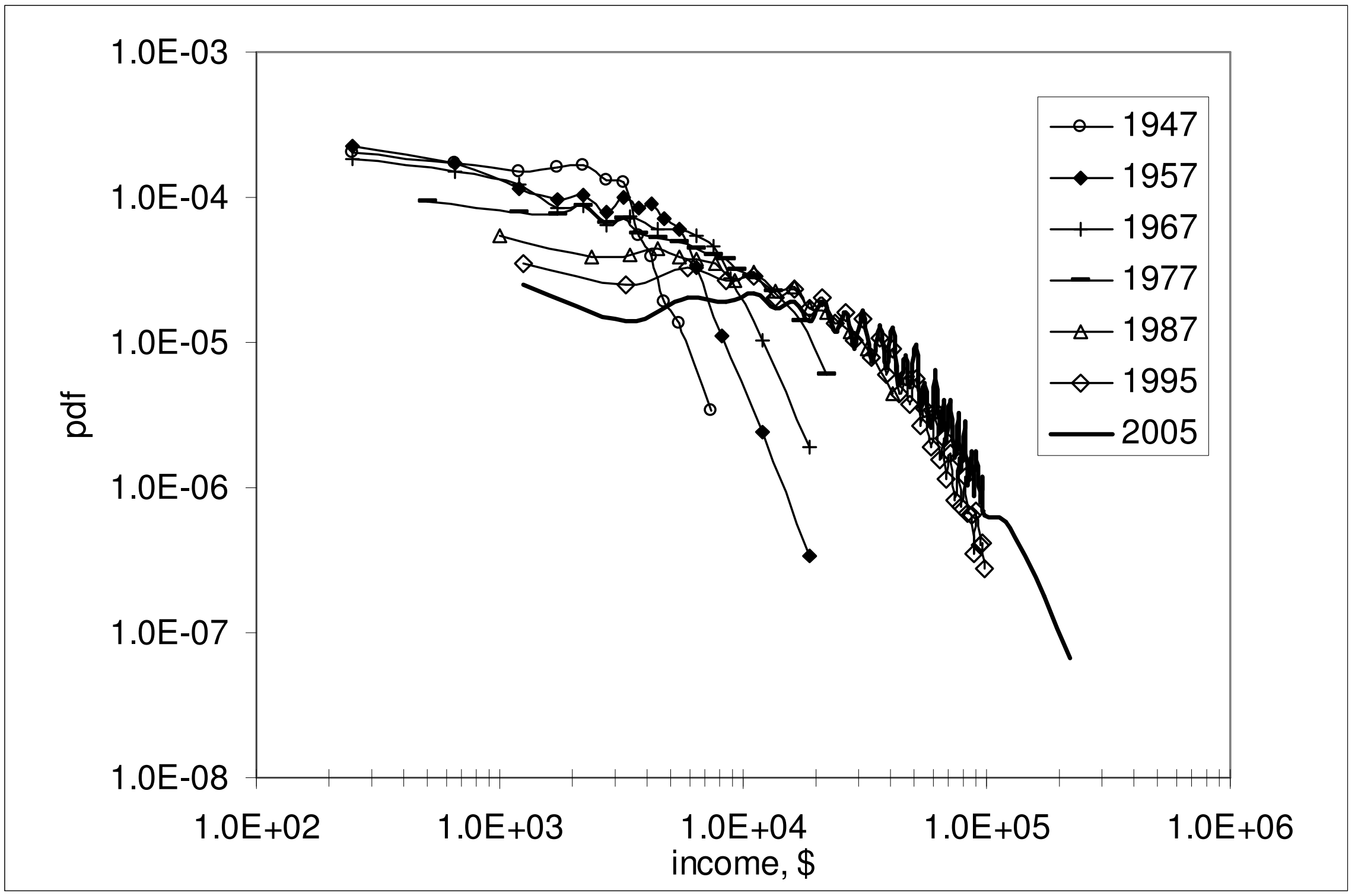

Figure 2. PIDs for selected years measured in current dollars: a) – for the years between 1947 and 1987 in constant income bins; b) – for the years between 1947 to 2005 in varying income bins. The PIDs are normalized to the total population and reduced to width of corresponding income bin.

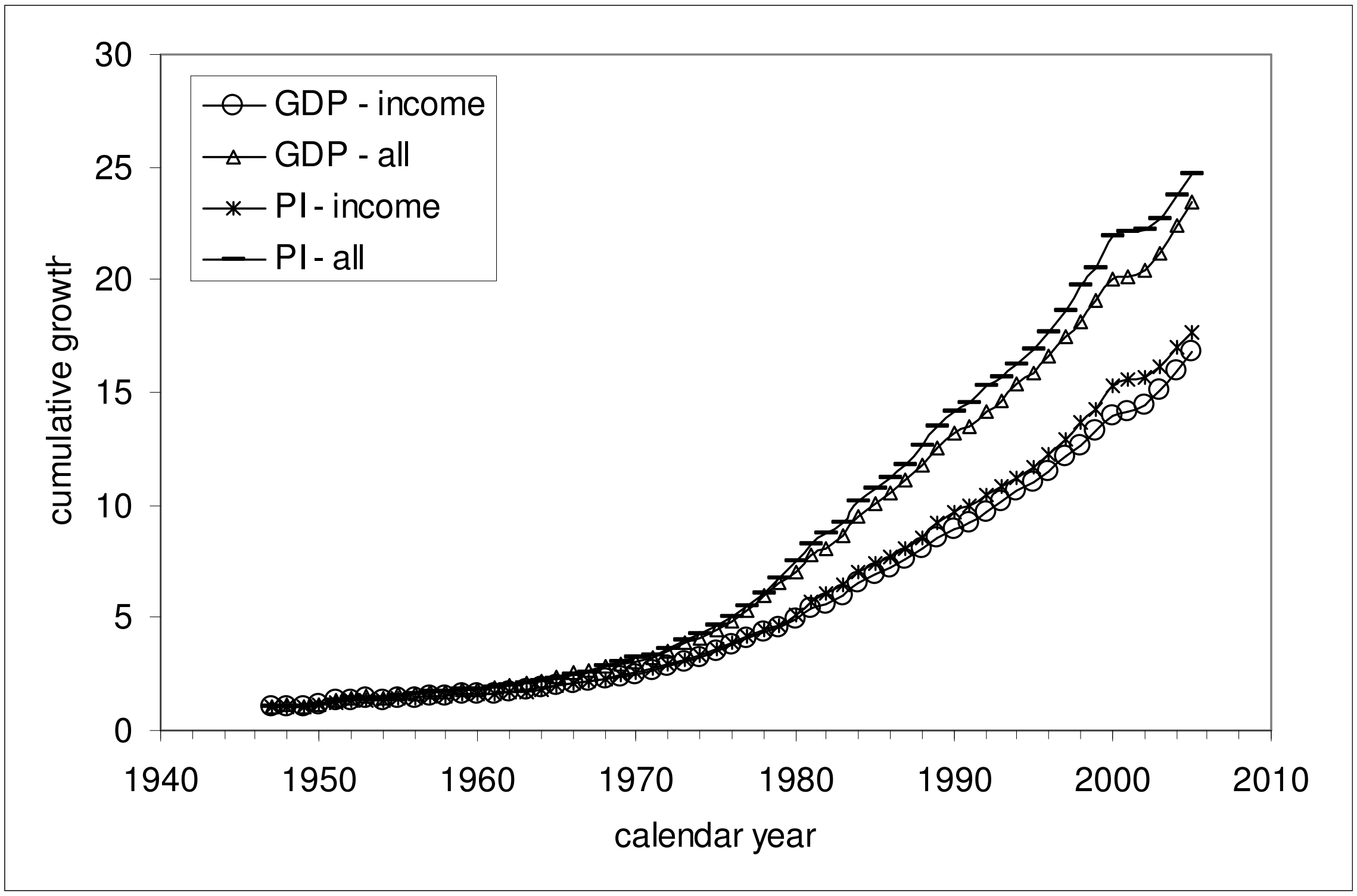

Figure 3. Cumulative growth of the nominal GDP per capita and nominal GPI per capita reduced to different population groups. The measured PIDs in Figure 4 are reduced to the GPI for the population with income.

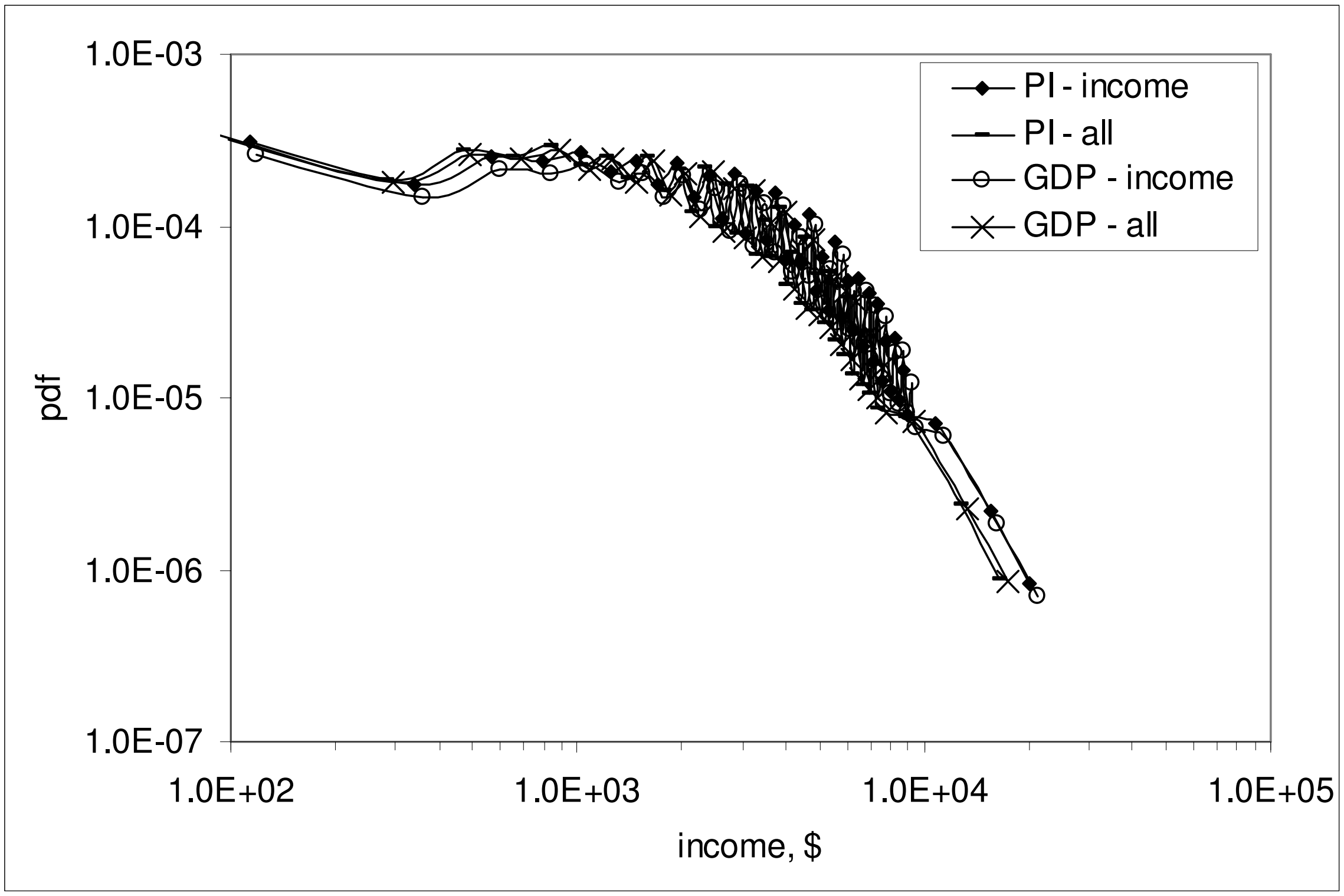

Figure 4. The PID for 2005 reduced to the cumulative growth between 1947 and 2005 of the four variables presented in Figure 3.

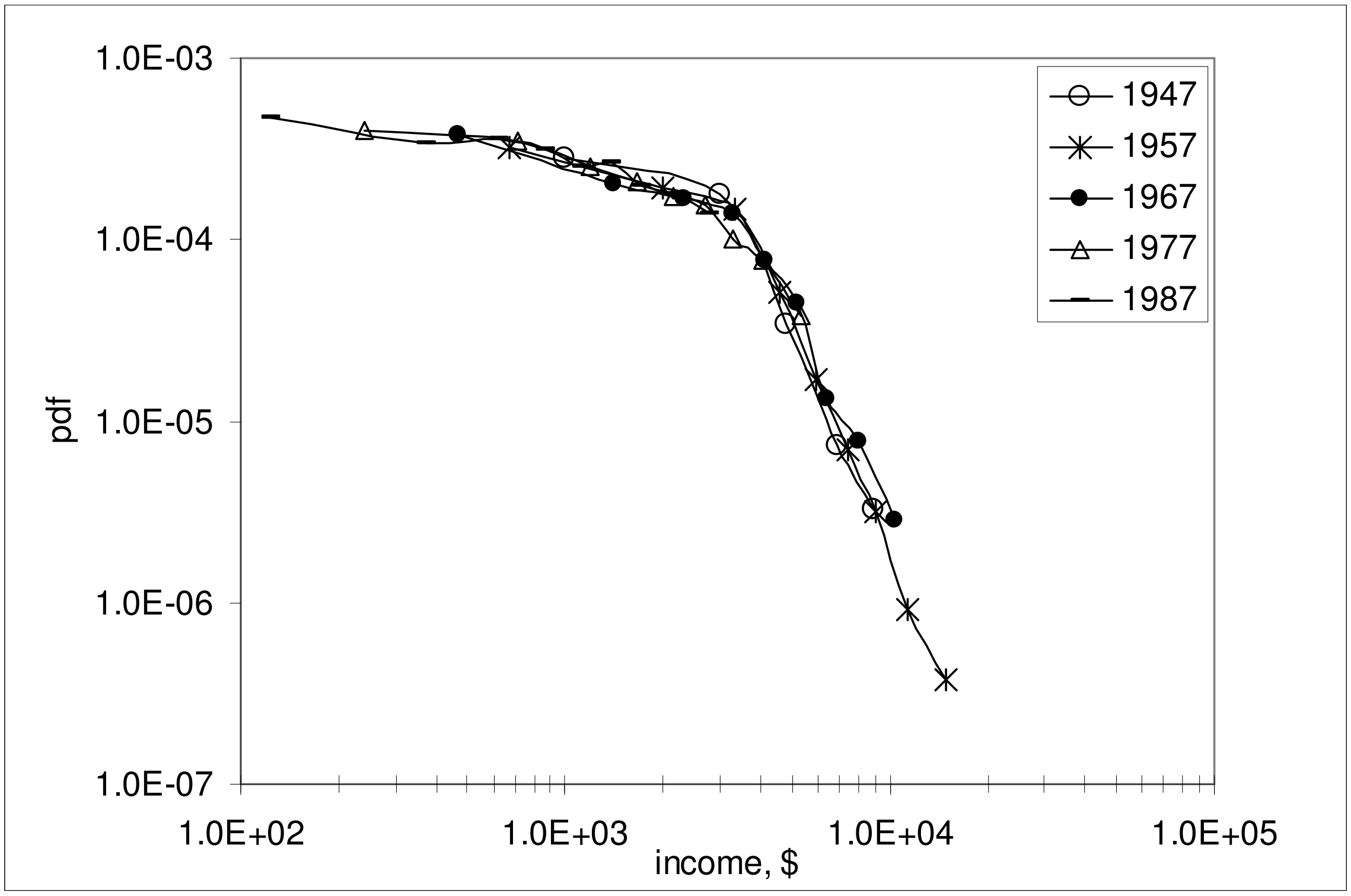

Figure 5. PIDs for some selected years between 1947 and 1987. The income scale is reduced by the cumulative growth of the nominal GPI per capita since 1947, as obtained for people with income. Notice consistent behavior of the PIDs between 1947 and 1987. One can expect an approximately constant true Gini coefficient for the years before 1987.

a)

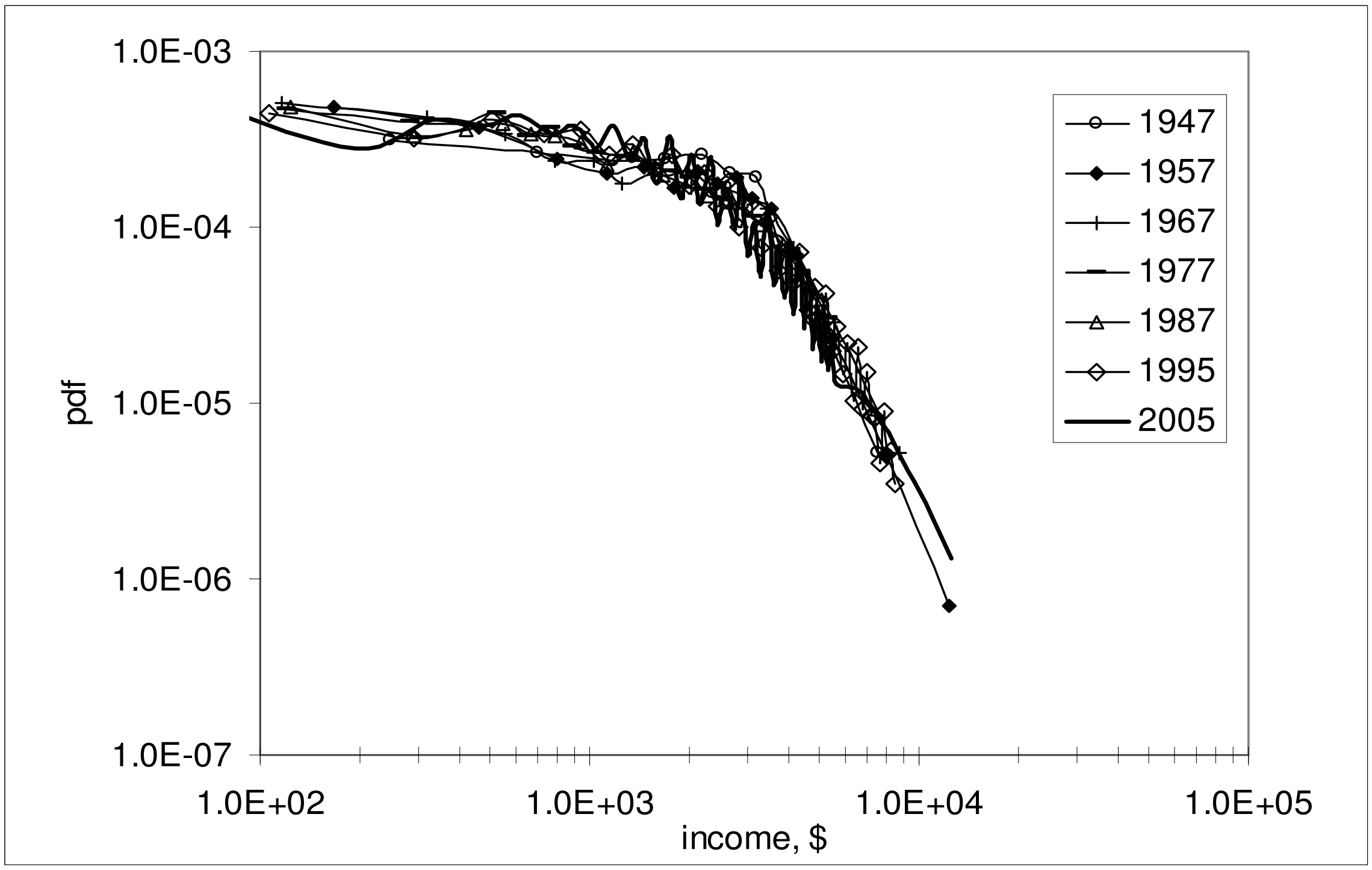

1995
2005

b)

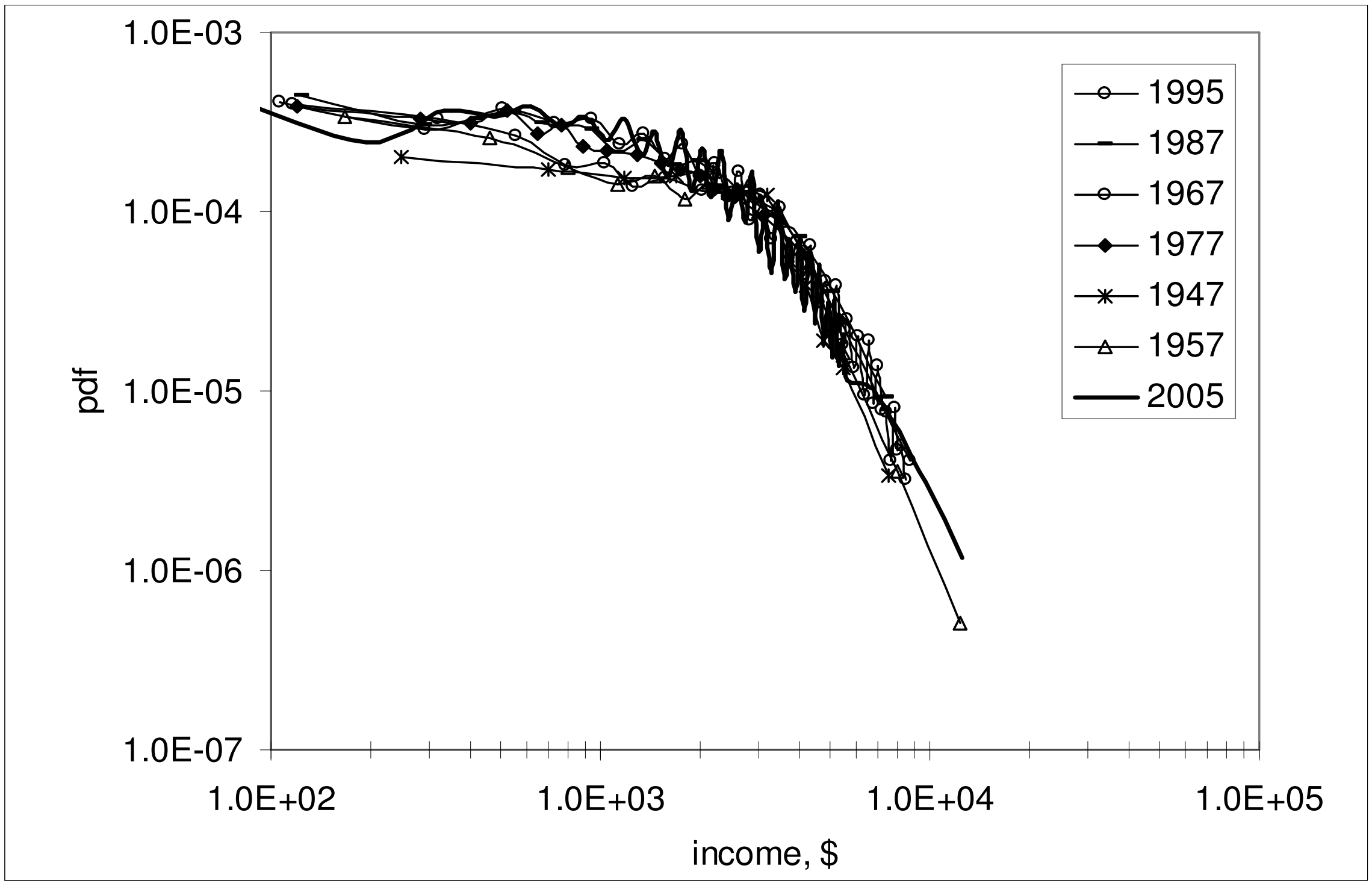

Figure 6. PIDs for some selected years between 1947 and 2005: a) – for people with income and b) – for all people 15 years of age and above. The income scale is reduced to the cumulative growth of the nominal GPI per capita since 1947, as obtained for people with income.

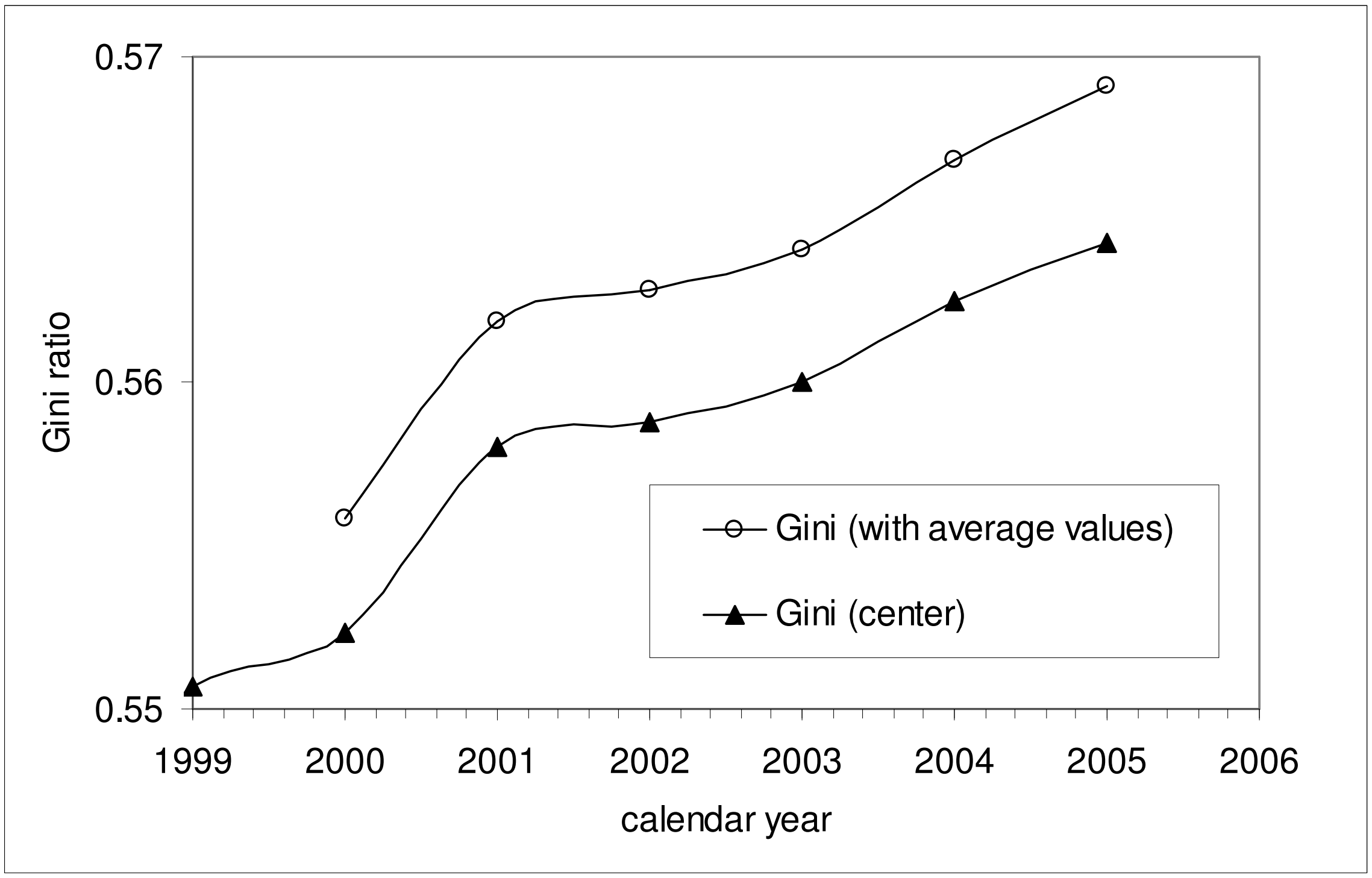

Figure 7. Comparison of Gini coefficients with and without usage of average income values in income bins. The curve using the average values also has three additional income bins between $100,000 and $250,000.

a)

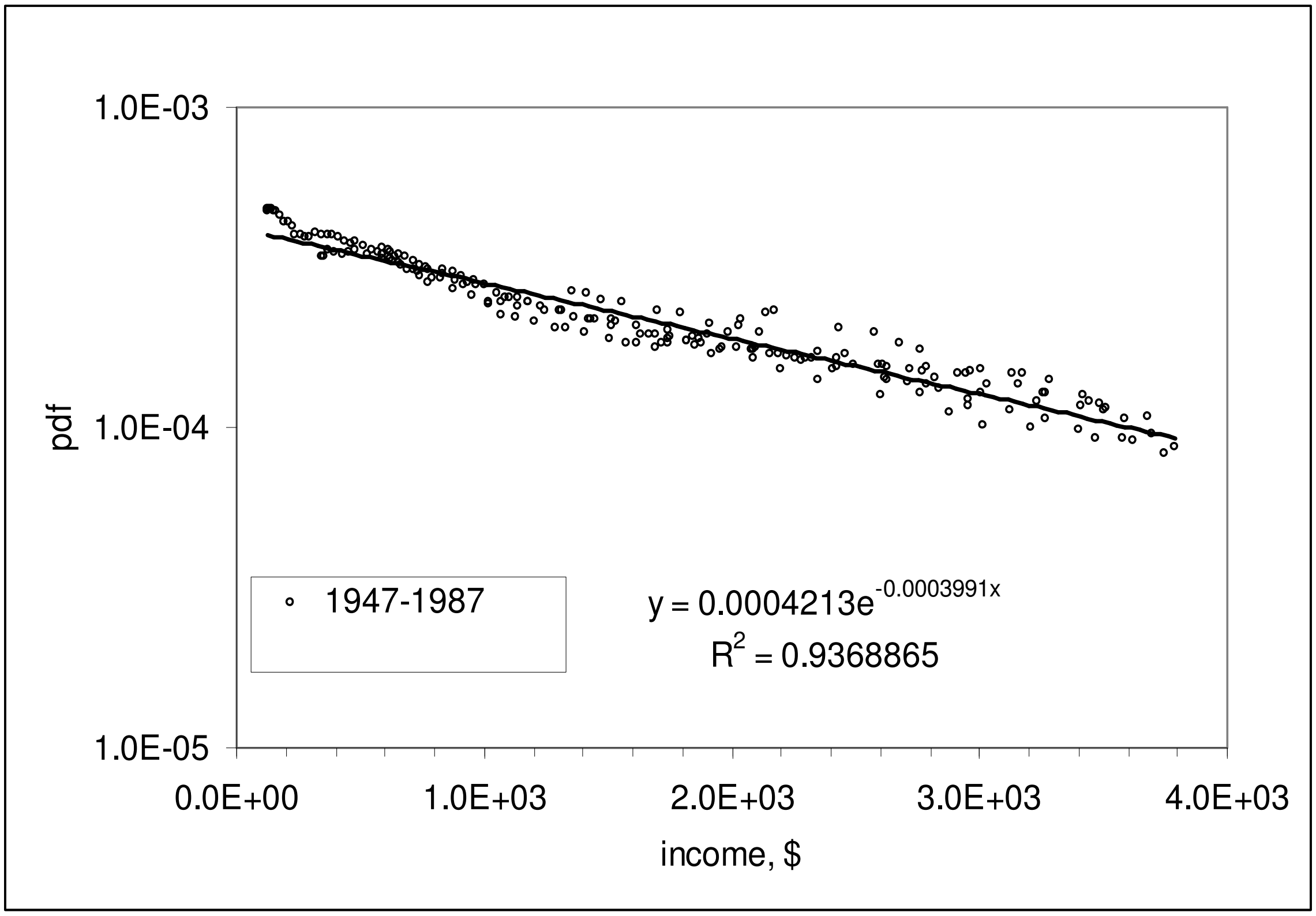

b)

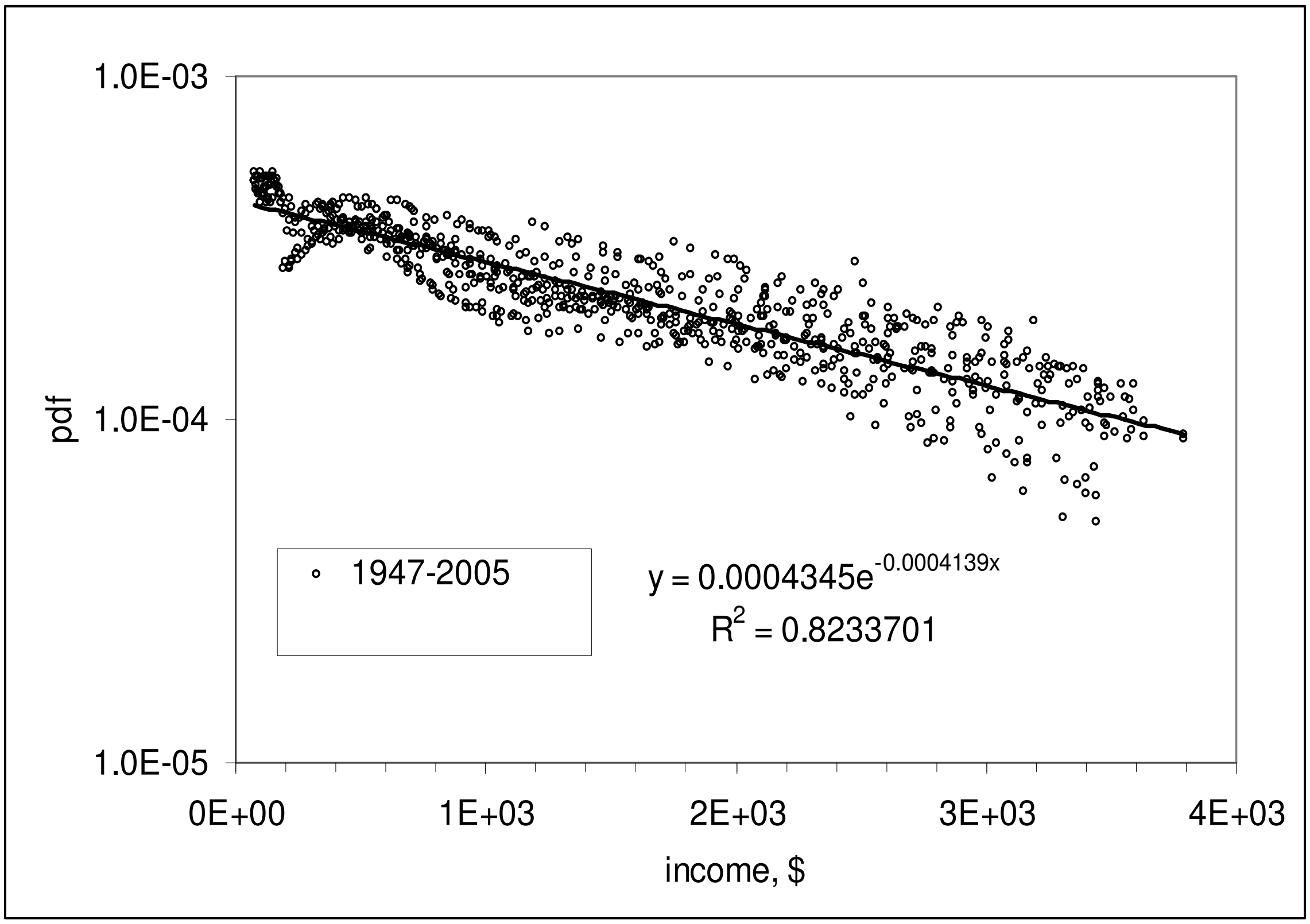

Figure 8. Approximation of the PIDs between 1947 and 1987 – a); and between 1947 and 2005 – b) by exponential functions. Obtained indices are very close, but scattering is large in the second case, which is also characterized by higher resolution.

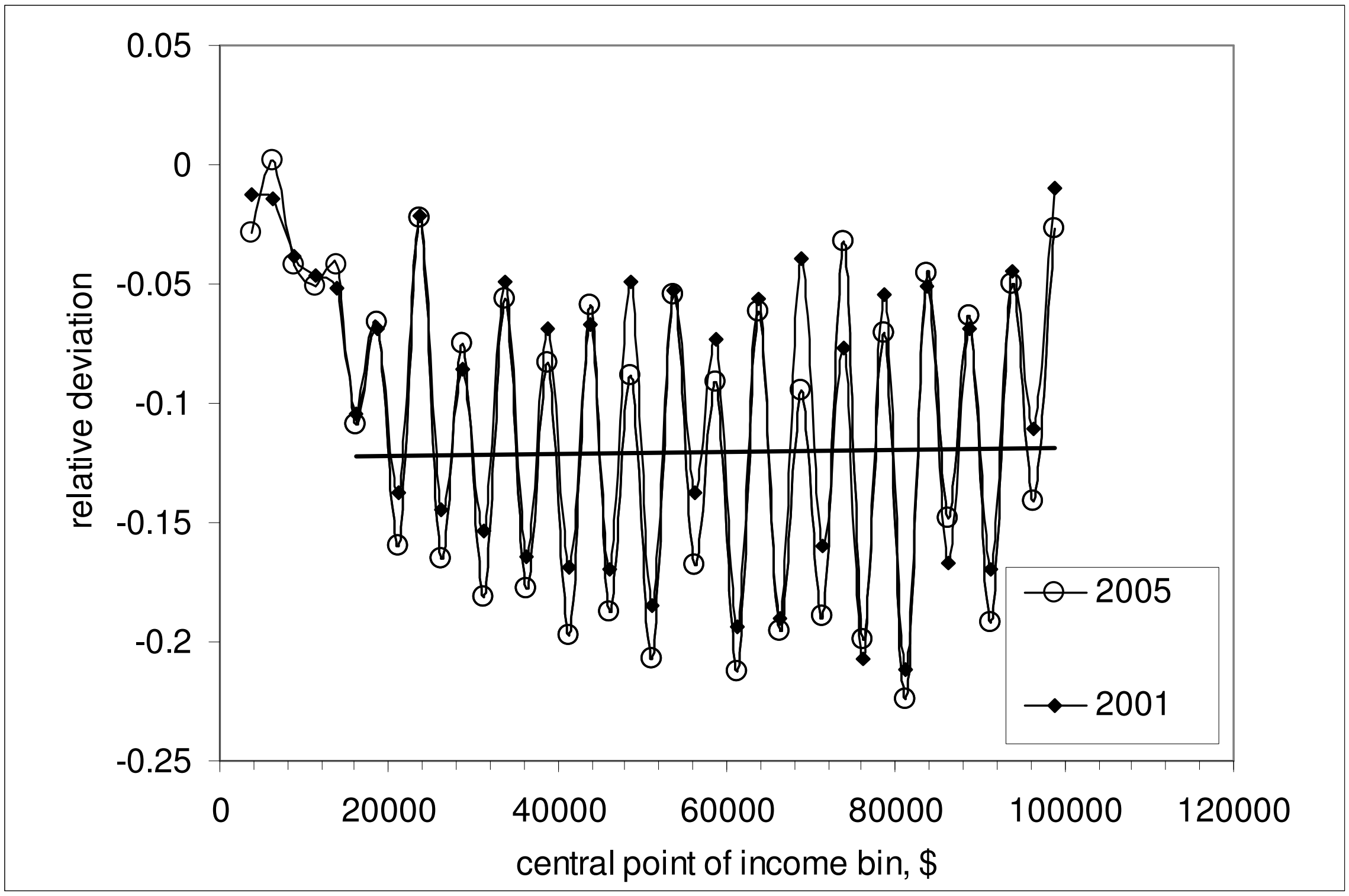

Figure 9. Relative deviation of the average value in income bins, $X_{av}$, from the central point of the bin, $X_c$ : $(X_{av}-X_c)/dX$, where $dX$ is the width of corresponding bin. The CPS reports for 2001 and 2005 are compared.

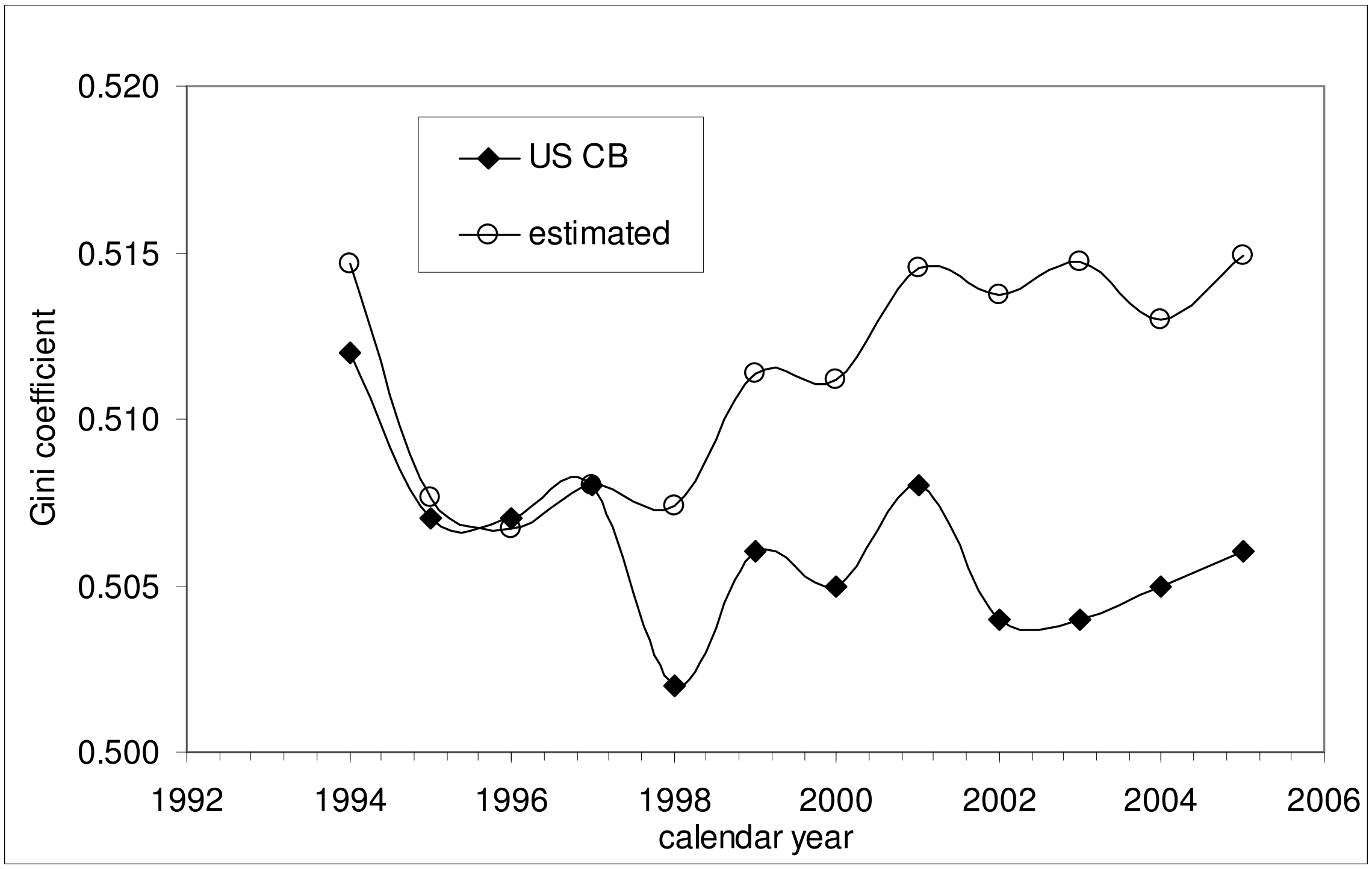

Figure 10. Comparison of the Gini coefficient reported by the US Census Bureau for the years between 1994 and 2005 with that estimated in this study. Obviously, the US CB changed the procedure for the computing of Gini coefficient in 1998.

a)

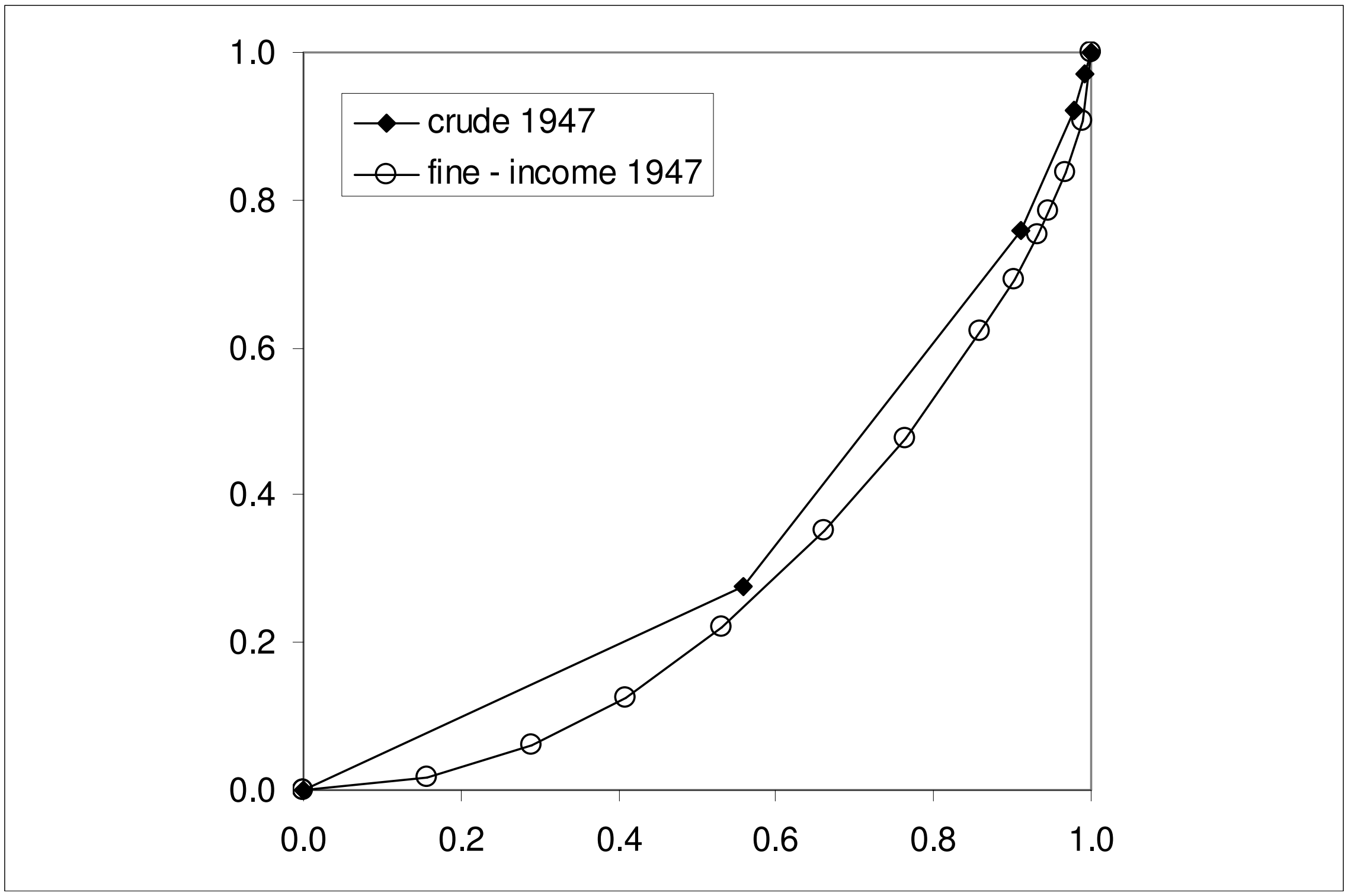

b)

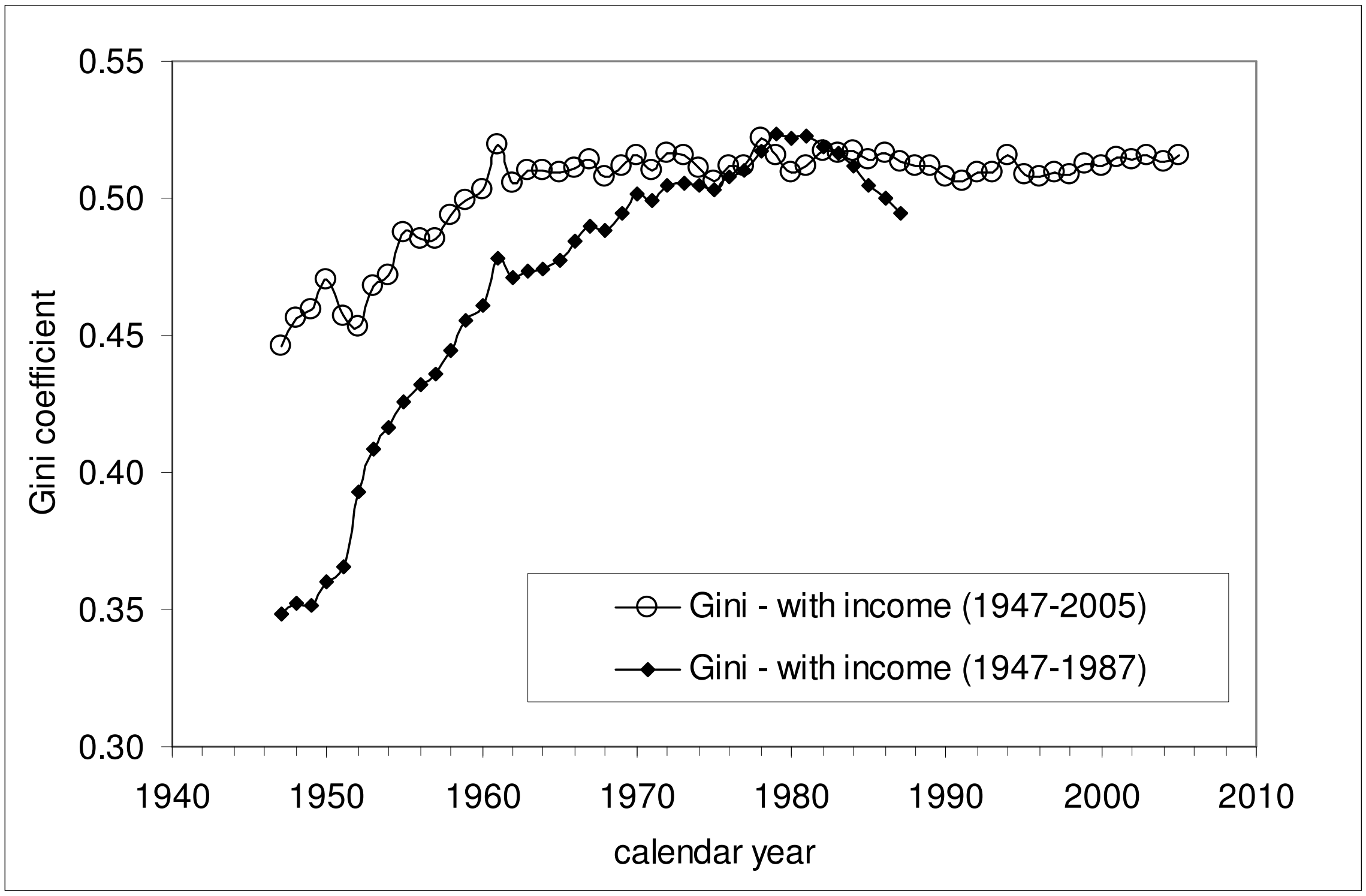

Figure 11. Comparison of two Lorenz curves for 1947 associated with the crude and fine PIDs –a). Comparison of two estimates of Gini coefficient between 1947 and 2005 using the crude and fine PIDs – b). Both coefficients are obtained for the population with income. The observed change in the actual PIDs is not well described by the fixed income bins. Nevertheless, the years between 1973 and 1983 are characterized by a good agreement between the curves.

a)

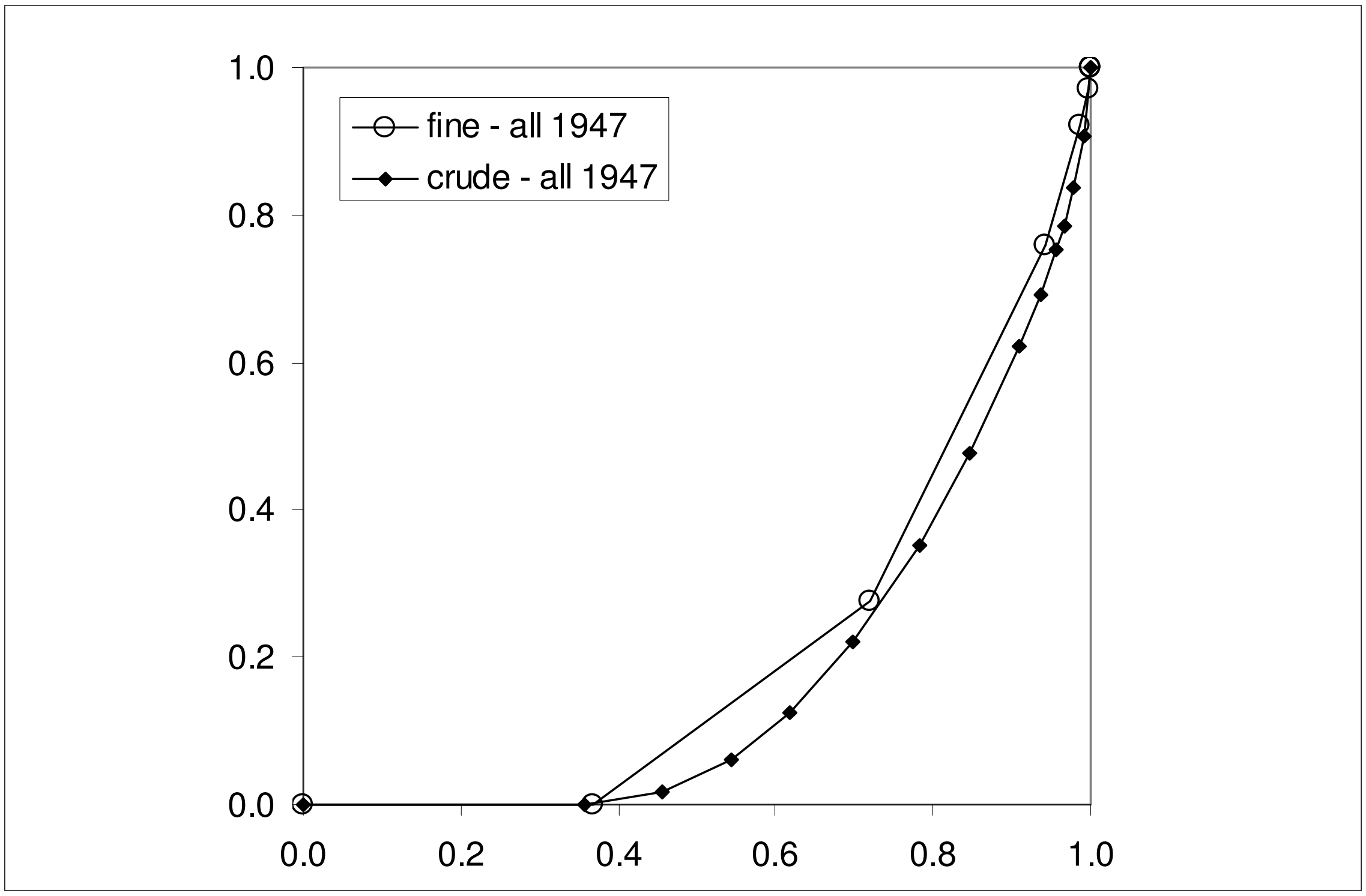

b)

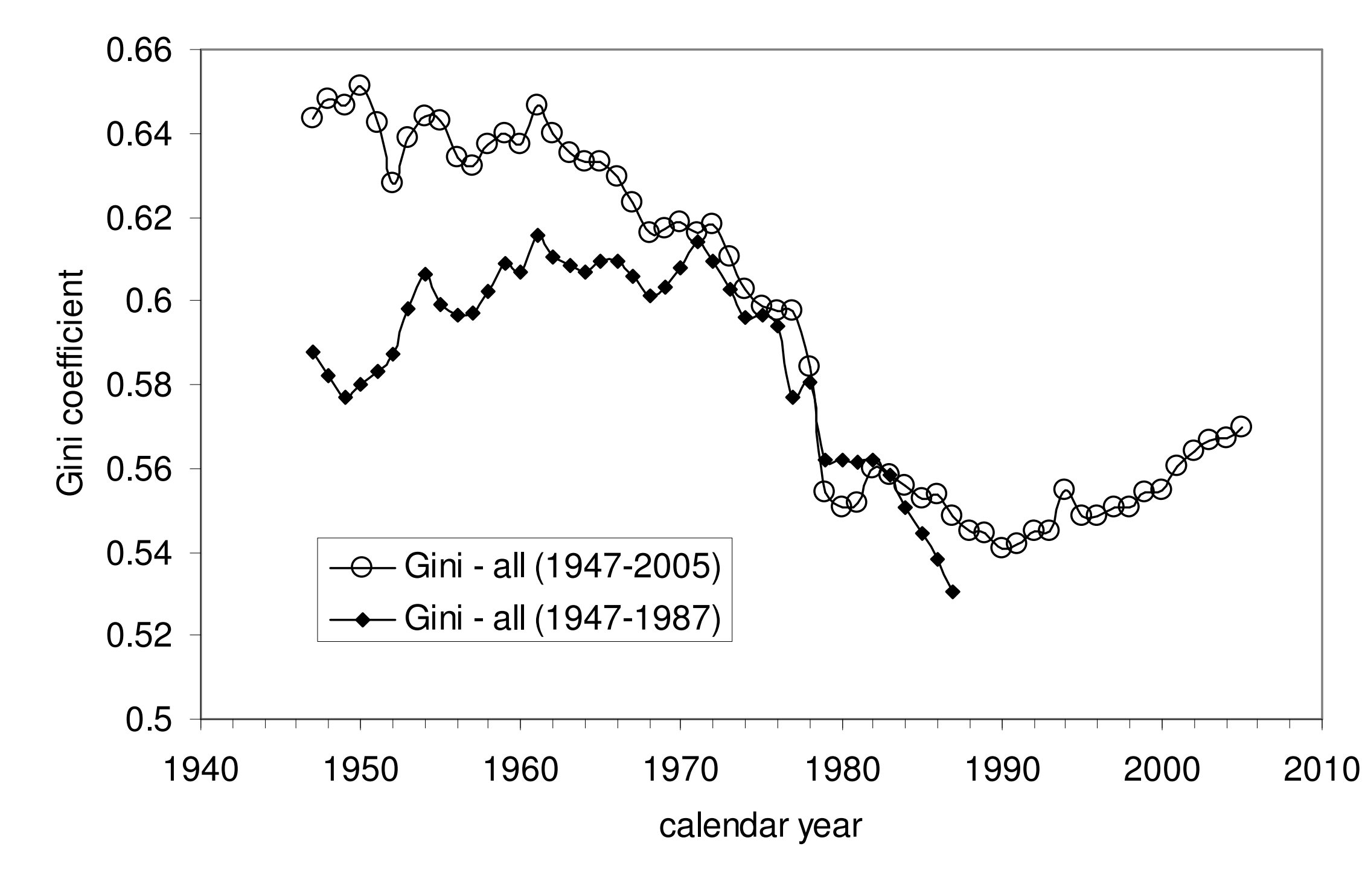

Figure 12. Comparison of two Lorenz curves for 1947 associated with the crude and fine PIDs –a). Comparison of two estimates of Gini coefficient between 1947 and 2005 using the crude and fine PIDs – b). Both coefficients are obtained for the whole working age population. The observed change in the actual PIDs is not well described by the fixed income bins. Nevertheless, the years between 1970 and 1983 are characterized by a good agreement between the curves.

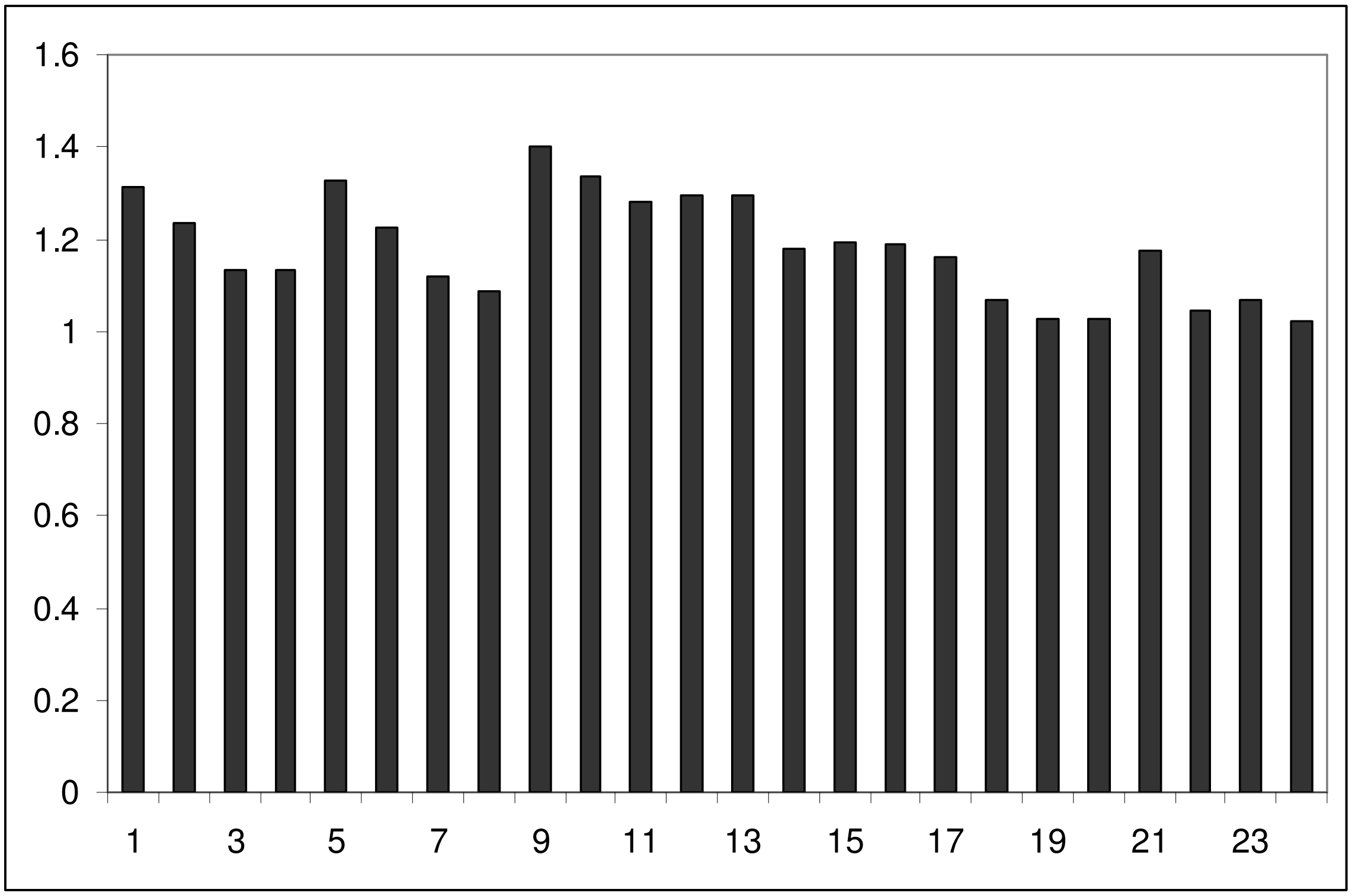

Figure 13. Estimates of the power law index *k* obtained from the average values of income in the Pareto income zone – above \$100000, \$150000, \$200000, and \$250000 for the years between 2000 and 2005.

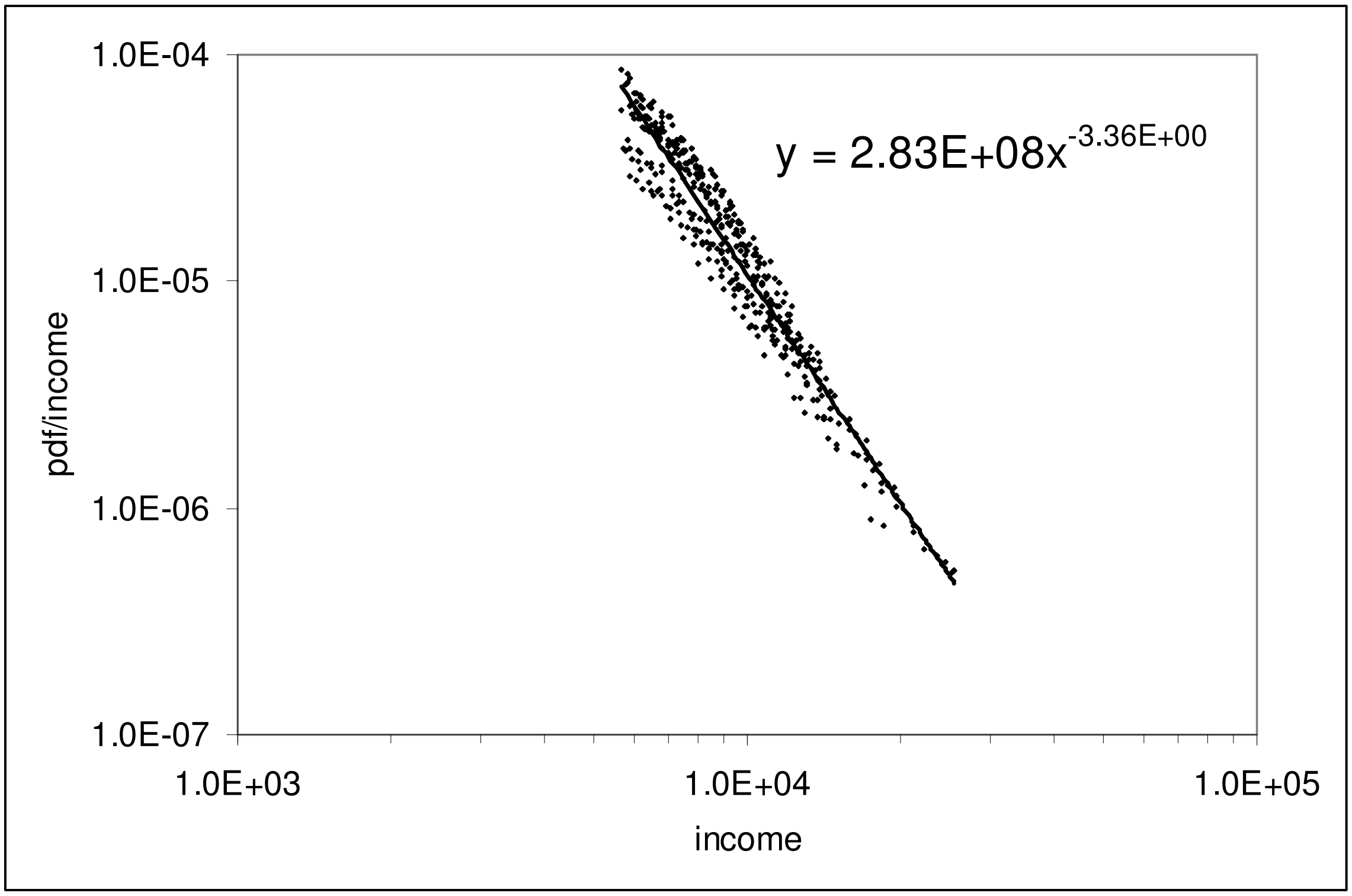

Figure 14. Linear regression of the normalized PIDs in the Pareto zone (log-log coordinates). The Pareto index is $k$=1.36. This estimate is consistent with that obtained using the average values above $100000 in Figure 13. Notice that the PIDs are divided by income bin widths in order to obtain density function independent on the widths. This procedure increases the power law index $k$ by one unit.

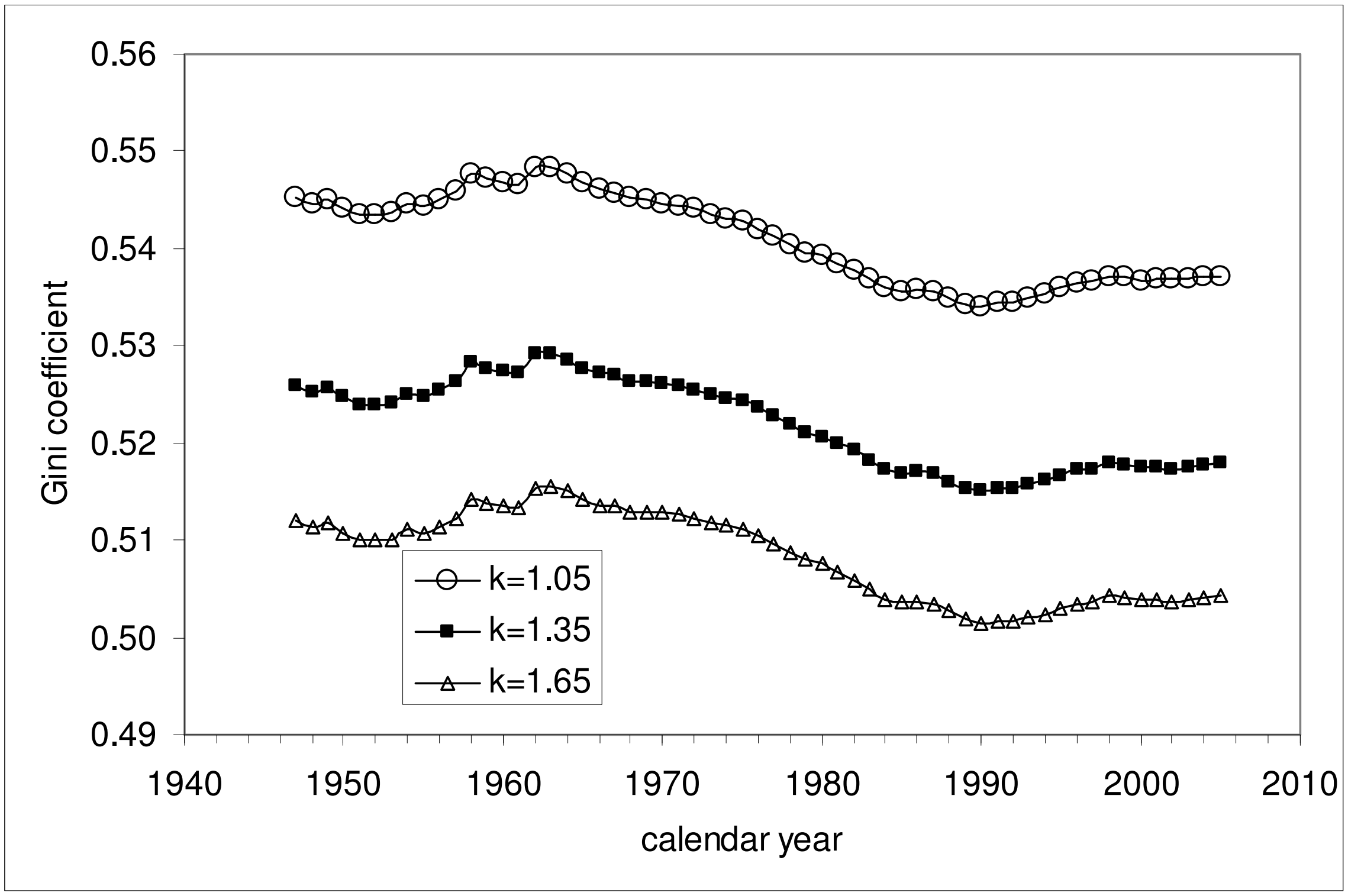

Figure 15. Dependence of predicted Gini coefficient on the Pareto index, *k*. Lower *k* values create "thicker" tails in PID and larger *G* values. The effect of *k* on *G* is nonlinear.

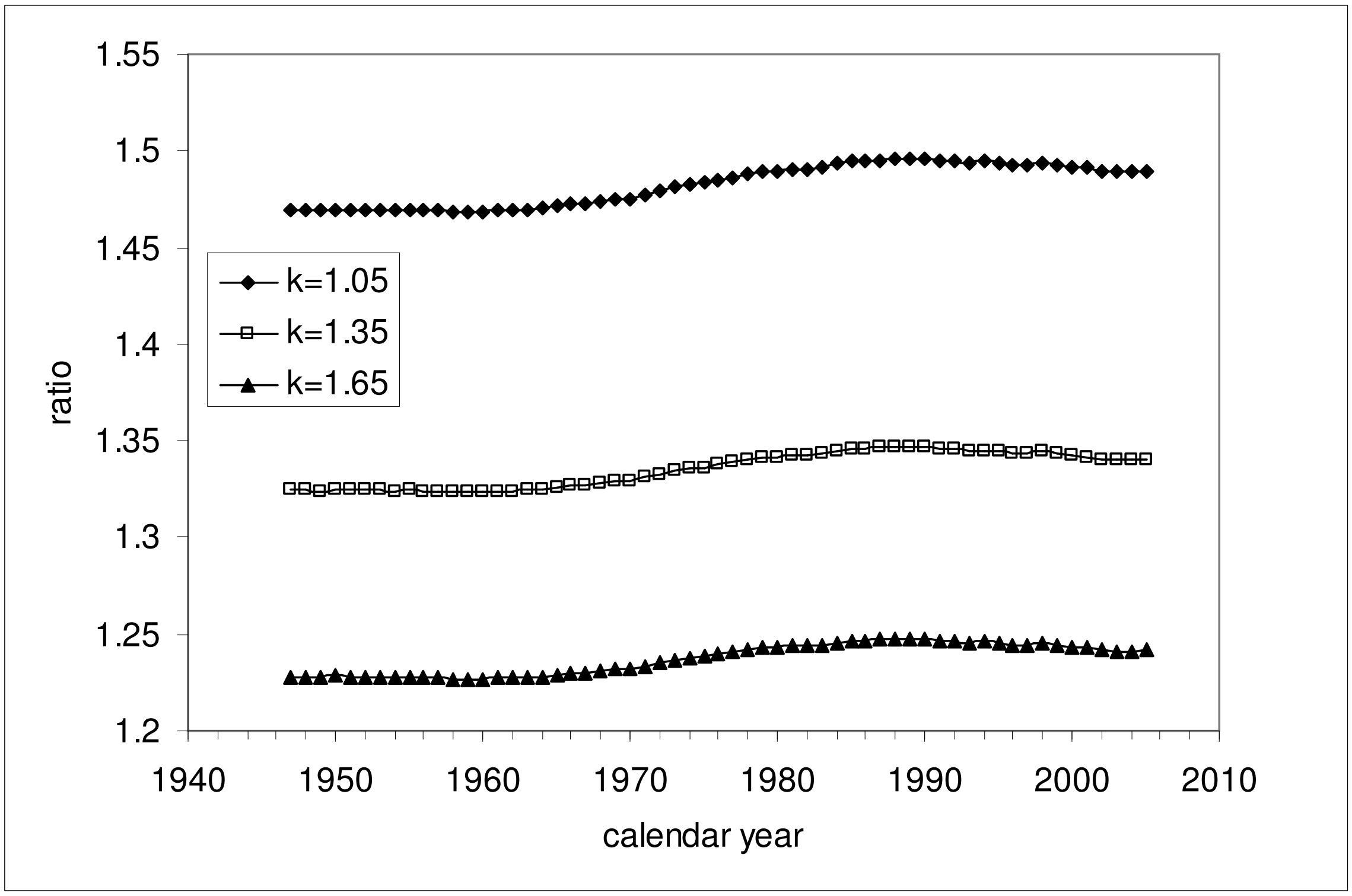

Figure 16. Dependence of the effective increase in income production (extra income) in the model relative to that in the sub-Pareto income zone (Kitov, 2005c). Theoretical value is 1.33 and corresponds to *k*=1.35. The effect of *k* on the ratio is nonlinear.

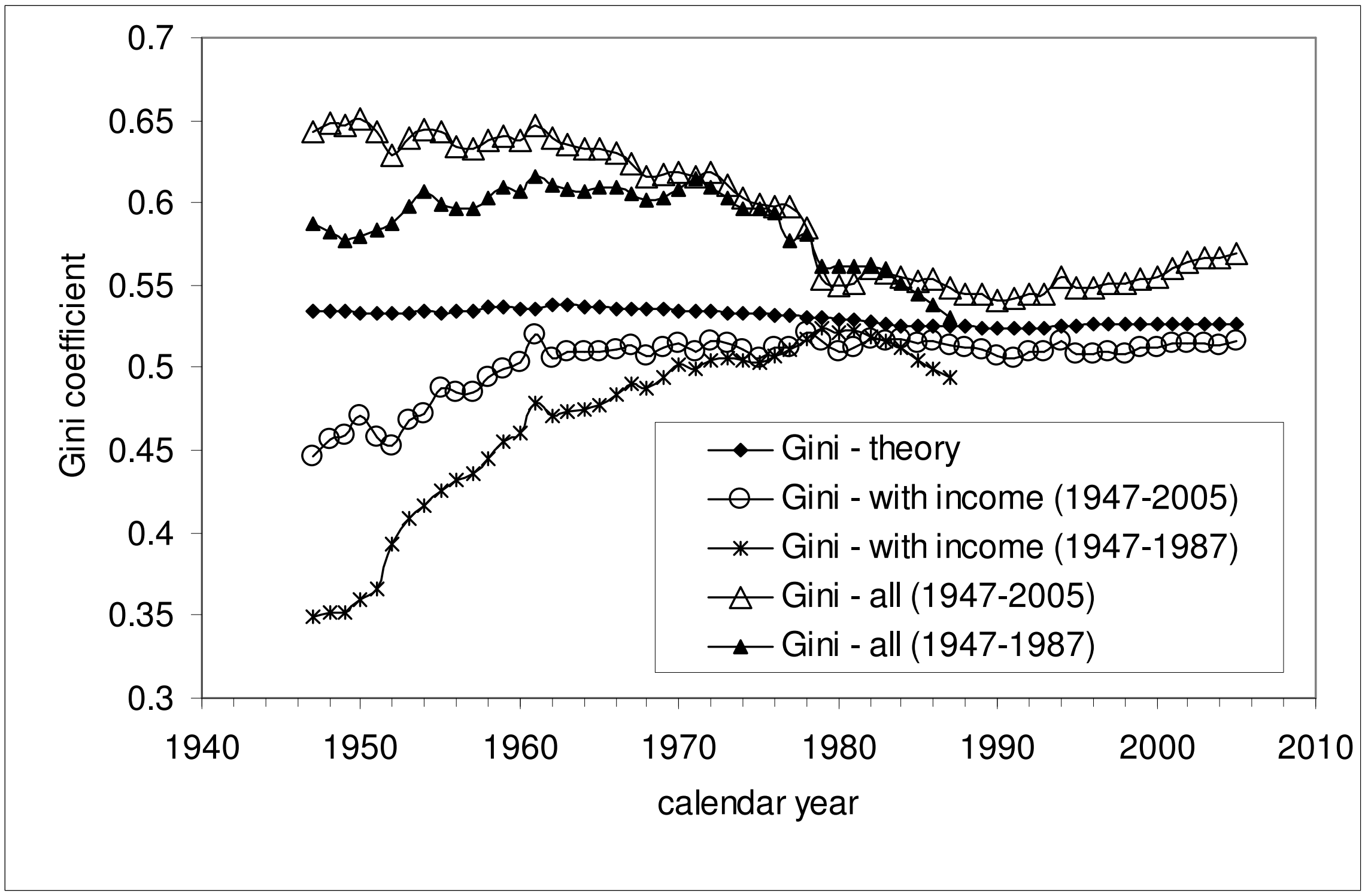

Figure 17. Comparison of the estimated and predicted Gini coefficients. The predicted curve lies between the two estimated curves, which converge as the portion of population without income drops. One can consider the predicted curve as representing the true Gini coefficient for the period between 1947 and 2005.